\documentclass[10pt]{article}
\usepackage{amsfonts,amsmath,amssymb}
\usepackage{graphicx}

\usepackage[makeindex]{imakeidx}
\makeindex

\usepackage{hyperref}
\hypersetup{
 colorlinks,
   citecolor=green,
   filecolor=green,
    linkcolor=blue,
 urlcolor=green
 }

\newfont{\twelvemsb}{msbm10 scaled\magstep1}
\newfont{\eightmsb}{msbm8}
\newfam\msbfam
\textfont\msbfam=\twelvemsb \scriptfont\msbfam=\eightmsb
\catcode`\@=11
\def\Bbb{\ifmmode\let\next\Bbb@\else
\def\next{\errmessage{Use \string\Bbb\space only in math mode}}\fi\next}
\def\Bbb@#1{{\fam\msbfam{{#1}}}}

\newcommand{\be}{\begin{equation}}
\newcommand{\ee}{\end{equation}}
\newcommand{\ba}{\begin{eqnarray}}
\newcommand{\ea}{\end{eqnarray}}
\newcommand{\q}{\theta}
\newcommand{\nn}{\nonumber}
\newcommand{\m}{\mathcal}
\newcommand{\we}{\varepsilon}

\begin{document}

\sloppy
\renewcommand{\thefootnote}{\fnsymbol{footnote}}
\newpage
\setcounter{page}{1} \vspace{0.7cm}
\vspace*{1cm}
\begin{center}
{\bf Strong Wilson polygons from the lodge of free and bound mesons} \\
\vspace{.6cm} {Alfredo Bonini$^a$, Davide Fioravanti $^a$, Simone Piscaglia $^{b}$, Marco Rossi $^c$}
\footnote{E-mail: bonini@bo.infn.it, fioravanti@bo.infn.it, piscagli@to.infn.it, rossi@cs.infn.it} \\
\vspace{.3cm} $^a$ {\em Sezione INFN di Bologna, Dipartimento di Fisica e Astronomia,
Universit\`a di Bologna} \\
{\em Via Irnerio 46, 40126 Bologna, Italy}\\
\vspace{.3cm} $^b$ {\em Dipartimento di Fisica and INFN, Universit\`a
di Torino} \\
{\em Via P. Giuria 1, Torino, Italy}\\
\vspace{.3cm} $^c${\em Dipartimento di Fisica dell'Universit\`a
della Calabria and INFN, Gruppo collegato di Cosenza} \\
{\em Arcavacata di Rende, Cosenza, Italy} \\
\end{center}
\renewcommand{\thefootnote}{\arabic{footnote}}
\setcounter{footnote}{0}
\begin{abstract}
{\noindent Previously predicted by the $S$-matrix bootstrap of the excitations over the GKP quantum vacuum, the appearance of a new particle at strong coupling --  formed by one fermion and one anti-fermion -- is here confirmed: this two-dimensional meson shows up, along with its infinite tower of bound states, while analysing the fermionic contributions to the Operator Product Expansion (collinear regime) of the Wilson null polygon loop. Moreover, its existence, free \footnote{This term is used here as opposite to bound, thus as {\it unbound}.} and bound, turns out to be a powerful idea in re-summing all the contributions (at large coupling) for a general $n$-gon ($n\geq 6$) to a Thermodynamic Bethe Ansatz, which is proven to be equivalent to the known one and suggests new structures for a special $Y$-system.}

\end{abstract}
\vspace{6cm}

\newpage


\tableofcontents

\index{}
\printindex
\newpage


\section{Introduction and purposes}\label{intro}
In the recent time, maximal supersymmetric quantum field theory ${\cal N}=4$ SYM (Super Yang-Mills) -- in the multi-colour limit $N_c\rightarrow \infty$ with 't Hooft coupling $N_c g_{YM}^2$ parametrised as $\lambda = 16 \pi ^2 g^2$ --  revealed itself as the stage of a large scientific progress of understanding: unveiling integrability \cite{MZ} and its developing ({\it cf.}, {\it e.g.} \cite{BS1,BS2,BS3,BS4,BS5,BES,TBA1,TBA2,TBA3,QSC1,QSC2}\footnote{The reader interested in condensed matter consequences of this rapid expansion might look at \cite{CCMT} and its references.}) has led to enormous progress towards a better comprehension and proof of its duality (at large coupling) with a gravity theory, the so-called AdS/CFT correspondence \cite{MGKPW1,MGKPW2,MGKPW3}. Although, these references concern mainly the spectrum of local operators, even more recently some approach based on integrability have been devised and pursued for the 4D scattering amplitudes or, which is the same \cite{AM-amp, DKS, BHT}, polygonal Wilson loops (WLs), namely operators among the simplest ones of non-local nature.

A specific integrability-based approach has proposed to expand (in the collinear limit) the expectation value of a null polygonal Wilson loop in a sort of Operator Product Expansion (OPE) \cite{TBuA,YSA,Anope}\footnote{For an alternative, physical interpretation in terms of wave functions of the open GKP spin chain and their scalar products, we wish to refer the reader to the intriguing approach of Belitsky \cite{Bel-ope}.}, very much resembling the Form Factor (FF) Infra-Red (IR) spectral series of two-point correlation function in integrable quantum field theories (for an overview on the literature, {\it cf.} the reviews \cite{mussardo} and \cite{smirnov-book}). More in an integrability perspective, this point of view is severely based on the dispersion laws of the excitations over the GKP \cite{GKP} string vacuum \cite{Basso} and the 2D scattering factors \cite{FRO6, FPR1, Basso-Rej}, exactly as in the aforementioned FF bootstrap approach ({\it e.g.} \cite{mussardo} and \cite{smirnov-book} and references therein). In fact, in the last two years a series of papers by Basso, Sever and Vieira (BSV) \cite{BSV1,BSV2,BSV3,BSV4,BSV5,BCCSV1,BCCSV2,BSV6} has been making more and more precise the series and its single terms, so that to reconstruct expectation values of null polygonal Wilson loops, especially at weak coupling. Actually, this series is perturbative in the exponential of the cross ratios (of the polygon)\footnote{These are the analogue of the adimensional renormalisation time $r=m R$ of the two-point FF series with $m$ the mass scale and $R$ the distance between the two fields.} in the collinear limit (IR limit $r\rightarrow \infty$ of FFs) of the sides, but each term could be in principle determined exactly in the coupling $\lambda$ (like the multi-particle FFs). Of course, very much interest is attracted by virtue of the fact that the very same results yield Maximum Helicity Violating (MHV) 4D gluon scattering amplitudes.

Naturally, for specific tests one has to expand {\it each term of the series} \footnote{The attentive reader may already notice the arising of order of limits issues: as for the strong coupling regime, this shall actually be the main core of the present work.} at either weak (see, for instance, \cite{BSV2,BSV6,P1,P2,DP} and references therein) or strong coupling \cite{BSV3,BSV4,FPR2}, and compare respectively against gauge or string theory computations. Of course there are also serious hints that this method is more efficient than both perturbative theories, and also shall give results in the unexplored intermediate region. Actually, string theory has so far given only the (classical) leading order (LO) at large $g$ by means of a rather sophisticated mathematical minimisation of the bubble insisting on the $AdS$ polygon \cite{TBuA,YSA,Anope, Hatsuda:2010cc}. In view of the one-loop corrections to the Nekrasov-Shatashvili \cite{NS} prepotential \cite{Bourgine2015a, Bourgine2015b} ({\it cf.} \cite{FPR2} for the similarity with the present series), we shall hope that this method should give the next-to-leading order (NLO) soon and {\it easily}; we will be glad to comment on this point in section \ref{conc}, {\it Conclusion and outlook}.

Already in the paper \cite{FPR2}, we have computed the single terms and re-summed, at (LO) large coupling, the OPE for the hexagon \cite{Anope}, as proposed by BSV \cite{BSV1}, but with an important modification, thanks also to our previous results on the scattering \cite{FPR1} over the GKP vacuum \cite{GKP}. In specific, we deliberately inverted the order of collinear limit (the series) with the large coupling one in a educated manner: in fact, motivated by a pole in the fermion-antifermion $S$-matrix at generic coupling, we have performed a bootstrap on the (asymptotic) Bethe Anstaz equations of \cite{FPR2} and thus shown the existence of a new mass $2$ particle in two dimensions ({\it cf.} the discussion below and \cite{AM, frolov-tseytlin, Basso, zarembo-zieme, BSV3}), a {\it meson}, along with its infinite bound states, {\it only at strong coupling}. Correspondingly we have modified the BSV series {\it at large coupling} by adding the contributions of these particles (whilst initially they were not in the spectrum).

This mass $2$ boson was originally noticed by \cite{AM, frolov-tseytlin} as classical $AdS_3$ transverse fluctuation of the $AdS_5\times S^5$ string world-sheet. In fact, several authors (\textit{e.g.} \cite{Anope,Basso,zarembo-zieme}) have supposed that this particle (allegedly formed by two mass $1$ fermions) may exit the physical domain at finite values of the coupling constant, or, in other words, become {\it virtual} . In the context of the BSV series, it was shown in \cite{BSV3} that, in the infinite coupling limit, a mass $2$ contribution arises from a pole in the state of a small fermion and a small antifermion (similar phenomena are also said to happen for more fermions).

Actually, the Bethe Ansatz equations of \cite{FPR2} are ideal to understand the nature of the alleged bound state, the meson, as it is represented, at {\it generic coupling}, by a string of a small fermion, a small anti-fermion and a triad of (massless) 'isospin carriers' (opportunely arranged so to give rise to an $SU(4)$ R-symmetry singlet):
\ba
1&=& e^{iRp_f(u_{f,1})} \frac{u_{f,1}-u_{\bar f,1}+2i}{u_{f,1}-u_{\bar f,1}-2i} \ S^{(ff)}(u_{f,1},u_{\bar f,1}) \cdot \dots \, , \nonumber \\
1&=& e^{iRp_f(u_{\bar f,1})} \frac{u_{\bar f,1}-u_{f,1}+2i}{u_{\bar f,1}-u_{f,1}-2i} \ S^{(ff)}(u_{\bar f,1},u_{f,1})  \cdot \dots
\label {pot-string} \, ,
\ea
where the fermions ($f$) or the anti-fermions (${\bar f}$) have scattering prefactor $S^{(ff)}(u,v)=S^{({\bar f} {\bar f})}(u,v)=S^{({\bar f} f)}(u,v)=S^{(f {\bar f})}(u,v)$ in terms of rapidities $u,\,v$. Yet, the momentum of the fermion or anti-fermion, respectively $p_f(u_{f,1})$ and $p_f(u_{\bar f,1})$, have the wrong sign of imaginary part which do not compensate, as $R\rightarrow +\infty$, the pole of the scattering factor to give something $\sim 1$ \cite{FPR2}. Nevertheless, without any further problem the meson enjoys the following bootstrap formula for the scattering of two of them \cite{FPR2}:
\ba\label {SMM-virtual}
S^{(MM)}(u,v) = \frac{u-v+i}{u-v-i} S^{(ff)}(u+i,v+i) S^{(ff)}(u-i,v+i)S^{(ff)}(u+i,v-i)S^{(ff)}(u-i,v-i) \,.\nn\\
\ea
Moreover, analogous arguments and formul{\ae} hold when binding two or more mesons. Eventually, all this compels these states to live in the unphysical domain of the $S$-matrix as long as the coupling is finite, namely, to be virtual particles. But, when it becomes infinite, interestingly in the classical string regime\footnote{In this context a prediction of this scalar was given by \cite{AM}.}, all these acquire the 'dignity' of genuine scalar particles in that the above reasoning on the momentum does not hold any more \cite{FPR2}.

Furthermore, we confirm here accurately that the fermion and anti-fermion do not contribute at the LO as {\it free} (or unbound) particles (they are somehow confined at this order, albeit they start contributing at NLO, {\it cf.} Conclusion and outlook \ref{conc}).
And also the scalars allow a subdominant contribution\footnote {In fact, for scalars the situation is more interesting. Nonperturbative dynamics of scalars on $S_5$ generates a contribution \cite {BSV4} proportional to $\sqrt {\lambda}$; being a purely quantum effect, this term is missed by the minimal area procedure of \cite{TBuA,YSA,Anope}, which is obtained in the framework of classical strings.}.
In this way, we have found the MHV six gluon amplitude/expectation value of null hexagonal Wilson loop in terms of a set of coupled saddle-point equations of Thermodynamic Bethe Ansatz form (TBA-like) \cite{FPR2}. These are equivalent to those of the string minimal area \cite{TBuA,YSA,Anope}, and have been supposed to generalise to general null polygon. In summary, we have shown evidence that the excitations of the GKP string which give contributions in the (LO) large coupling regime are:

\medskip

$\bullet$ gluons (with two polarisations) and their bound states;

$\bullet$ mesons and their bound states.

\medskip

\noindent
At generic values of the coupling constant, the bound states of gluons enjoy a standard behaviour, whereas the meson \cite{BSV3,FPR2} and their bound states \cite{FPR2} live in the wrong domain, so that they are virtual. But the meson acquires the {\it status} of (real) particle in the infinite coupling limit \cite{BSV3,FPR2}, and on the same footing its infinite bound states \cite{FPR2}. Besides, we may say, {\it a posteriori} \footnote{We could also say {\it ex juvantibus} in the sense that we have cured or corrected the spectrum on which the BSV series, {\it plainly assumed}, runs.} that the assumption of their existence perfectly reproduces the string TBA of the hexagon.

These results call for a better mathematical justification and understanding of the fermionic contributions at large coupling\footnote{Apparently, this need was expressed in \cite{BSV3} as well.}, within the BSV series only, without $S$-matrix bootstrap. We have promised this analysis in \cite{FPR2}, and in the present paper we want to fulfil the promise: we will show that the strong coupling limit of the contribution from the (small) fermions and anti-fermions in the hexagon BSV series is indeed equivalent to our modified series with the presence of a mass $2$ meson and of all its $n$ bound states of mass $2n$, $n=1, 2, 3,\dots$. In addition, relying on the effective presence of mesons and their bound states, we extend to a general polygon the re-summation at strong coupling of the ({\it modified}) BSV series as in Section 11 of \cite{FPR2} as for the hexagon. Up to some subtleties on the integrations contours ({\it cf.} below), we can reproduce the minimal area results \cite{YSA,Anope}.

The paper is organised as follows. In Section \ref {birth} we show how, in the BSV series over the spectrum, the $1$-fermion plus $1$-antifermion state gives rise to a (string) classical contribution which can be interpreted as that of a mass $2$ state in 2D, the so-called meson. Then, we study a state made up of two fermions and two anti-fermions, and prove that it reproduces two classical terms: one is interpreted as a two unbound mesons state, the other as a single particle of mass $4$, arising as a bound state of two mesons. Now, having better understood the mechanism of formation of meson and meson bound states within the BSV series, we generalise to a general polygon the methodology of \cite{FPR2} which re-sums the series with a {\it modified} spectrum, including all these particles (the aforementioned pattern of excitations in the strong coupling regime). This is the topic of Section \ref {predictions} where we first work out the heptagon, then the general case. And this is preceded by Section \ref {class-str} where we summarise the TBA description of Wilson loop expectation values as in \cite{TBuA,YSA,Anope} and put it in a form suitable for comparison with our outcomes of Section \ref {predictions}. This detailed comparison of the two approaches happens in Section \ref {comparison}: we find agreement, after a simple redefinition of pseudoenergies (and some better specification of the integration contours which inevitably makes the final expressions more involved). In conclusion, this agreement is a fundamental signal in favour of the presence of the meson and its bound states as excitations of the {\it classical} GKP string. As for the integrability properties, we find useful in Section \ref {Y-sys} to derive new forms of the elegant functional equations, the so-called Y-system, corresponding to the TBA equations, and realising a sort of discretised integrable dynamics in the rapidity space. At the very end, in two appendices we collect many useful formul{\ae} which facilitate reproducing the computations in the main text.

\section{The birth of a meson (and its bound states)}
\label{birth}

\begin{figure} [htbp]
\centering
\includegraphics[width=0.85\textwidth]{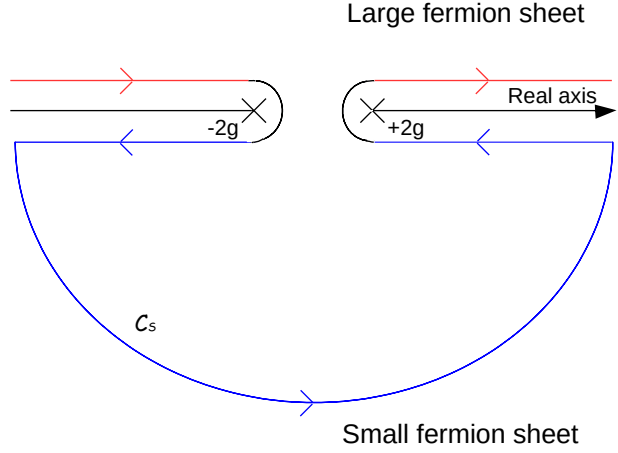}
\caption{The integrations on fermionic rapidities are performed along the path depicted in the figure above, going through the branch cut $[-2g,+2g]$; at strong coupling, the contribution due to large fermions (corresponding to the red section) is negligible, with respect to small fermion contribution. Therefore, throughout this section the integrations will be restricted to the path $\m{C}_{S}$, corresponding to the blue section of the curve, completely lying in the small fermion sheet.}
\label{Csmall}
\end{figure}

In \cite{FPR2}, as well as in the subsequent sections, for re-summing the BSV series of the hexagonal Wilson loop, an educated guess is assumed, namely that at LO strong coupling the fermionic excitations contribute only via their bound states, while the unbound fermion contributions are suppressed \cite{BSV3,FPR2}. In what follows, we are going to give evidence in favour of this claim, by showing how a couple of a small fermion and a small anti-fermion at large coupling forms a bound state, a so-called 'meson'; subsequently, a general picture is provided to enlighten the formation of mesons and bound states thereof, out of couples of fermions and anti-fermions.\\
In general, the hexagonal Wilson loop at generic coupling receives contributions from the intermediate state of all particles: scalars, gluons and fermions. With restriction on the latter fermionic sector, the contribution comes from $n$ small fermions plus $n$ small anti-fermions\footnote{A perfectly analogous contribution would come from the large fermions, but it is exponentially suppressed at strong coupling; we refer to \cite{Basso,FPR1,BSV2,FPR2} for exhaustive explanations. Therefore, throughout all this section the integrations are performed on a path $\m{C}_{S}$ in the small fermion sheet, as portrayed in Figure \ref{Csmall}; analogous path can be devised for large fermions.} in the form
\ba\label{WLferm}
&& \mathcal{W}_{hex}^{({\{f\} \{ \bar f\}})}=\frac{1}{n! n!}\int _{\m{C}_{S}}\prod_{k=1}^n \left[\frac{du_k}{2\pi}\frac{dv_k}{2\pi}
\,\mu_f(u_k)\mu_f(v_k)\,e^{-\tau E_f(u_k)+i\sigma p_f(u_k)+im_{f,k}\phi}\times \right.\\
&& \left. \times e^{-\tau E_f(v_k)+i\sigma p_f(v_k)+im_{\bar f,k}\phi}\right]
P(0|u_1\dots u_n,v_1\dots v_n)P(-v_n\dots -v_1,-u_n\dots -u_1|0)\nn \, ,
\ea
where the pentagonal transitions in the last line are factorised into a product of a dynamical and a matrix part \cite{BSV4,BCCSV1,BCCSV2}
\be\label{split}
P(0|u_1\dots u_n,v_1\dots v_n)P(-v_n\dots -v_1,-u_n\dots -u_1|0)=
\Pi_{dyn}^{(n)}(\{u_i\},\{v_j\})\,\Pi_{mat}^{(n)}(\{u_i\},\{v_j\}) \quad .
\ee
As also proposed by \cite{BSV6}, we write here an expression for the matrix part for $n$ fermions ($u_i$) and $n$ antifermions ($v_i$),
which reflects the $SU(4)$ $R$-symmetry, as encoded in the $Y$-system depicted in \cite{FFPT},
\ba\label{Pi_mat^n}
\Pi_{mat}^{(n)}(\{u_i\},\{v_j\})=\frac{1}{n!n!n!}\int_{-\infty}^{+\infty} \prod_{k=1}^n\left(\frac{da_k db_k dc_k}{(2\pi)^3}\right)
\,\frac{\displaystyle\prod_{i<j}^n g(a_i-a_j) g(b_i-b_j) g(c_i-c_j)}
{\displaystyle\prod_{i,j}^n f(a_i-b_j) f(c_i-b_j) \prod_{i,j}^n f(u_i-a_j) f(v_i-c_j)} \nn \, , \\
\,
\ea
where $f(x) \equiv x^2+\frac{1}{4} $ and $g(x) \equiv  x^2(x^2+1) $. Of course, this procedure, as recalls the view on Dynkin diagrams of \cite{FFPT}, should be fully general ({\it cf.} also the proposal of \cite{BSV4} for the scalars) and deserves further investigation.

For a couple of one  fermion and one anti-fermion, formula (\ref{Pi_mat^n}) reduces to a very simple shape (upon use of \cite{GradRyzh}):
\ba\label{mat1}
\Pi_{mat}^{(1)}(u,v) &=& \int_{-\infty}^{+\infty} \frac{da \,  db \, dc}{(2\pi)^3}
\,\frac{1}{(u-a)^2+\frac{1}{4}} \,  \frac{1}{(v-c)^2+\frac{1}{4}} \,
\frac{1}{(a-b)^2+\frac{1}{4}} \, \frac {1}{ (c-b)^2+\frac{1}{4}}= \nn\\
&=& \frac{4}{(u-v)^2+4} \ ;
\ea
on the other hand, the dynamical factor of (\ref{split}) for $n=1$ enjoys the form
\be
\Pi_{dyn}^{(1)}(u,v) = \frac{1}{P_{f\bar f}(u|v) P_{\bar f f}(v|u)} \ .
\ee
Taking into account both the dynamical and the matrix factors, a small fermion-antifermion couple contributes to the hexagonal Wilson loop with the amount
\ba\label{fer-antifer}
\mathcal{W}_{hex}^{(f\bar f)}&=&\int _{\m{C}_{S}}
\frac{du}{2\pi} \int  _{\m{C}_{S}} \frac{dv}{2\pi}\mu _f(u) \mu _{f}(v) e^{-\tau E_f(u) +i\sigma p_f (u)}
 e^{-\tau E_f(v) +i\sigma p_f (v)} \times \\
&\times & \frac{4}{(u-v)^2+4} \ \frac{1}{P_{f\bar f}(u|v)P_{\bar f f}(v|u)} \nonumber \, .
\ea

\begin{figure} [htbp]
\centering
\includegraphics[width=0.85\textwidth]{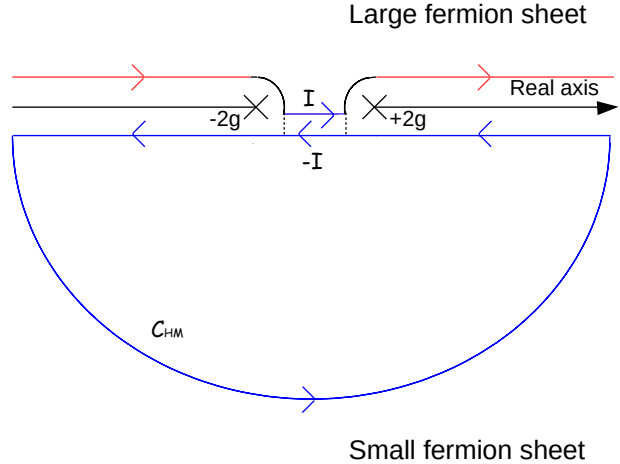}
\caption{Adding and subtracting the open interval $\m{I}=[-2g,+2g]$ depicted in figure, the integration contour $\m{C}_{S}$ can be seen as a sum of a closed curve $\m{C}_{HM}$, with half-moon shape, and an interval $\m{I}$ in the small sheet just below the branch cut (which involves unphysical values of the rapidity).
}
\label{Cfigura2}
\end{figure}

\begin{figure} [htbp]
\centering
\includegraphics[width=0.85\textwidth]{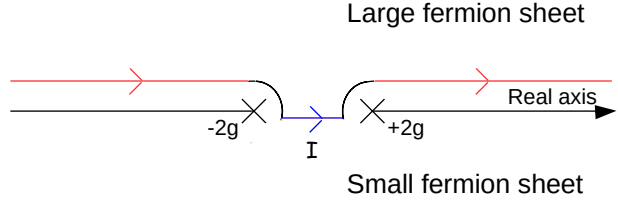}
\caption{Since no pole is any longer present in the lower half plane, the path $\m{C}_{S}$ depicted in Figure \ref{Csmall} can be deformed with continuity by shrinking the half circumference to the straight segment $\m{I}$, right below the branch cut. This is equivalent to say that we have no contribution from $\m{C}_{HM}$.}
\label{CsmallShrinks}
\end{figure}

Before performing the integration over the variable $v$, it is better to write our contour $\m{C}_{S}$ in two parts, $\m{C}_{HM}$ and $\m{I}=[-2g,+2g]$, according to Figure \ref{Cfigura2}, to highlight the different contributions. Now, we suppose on physical and mathematical grounds that in the small sheet $P_{f\bar{f}}(u|v)$ is free from zeroes and poles, while $P_{ff}(u|v)$ is free of zeros and its only pole is the one for coinciding rapidities, which gives rise to the existence of the fermion itself ({\it cf.} Appendix \ref{allcouplings} and \cite{BSV3}). Then, the integration of $v$ on $\m{C}_{HM}$ is evaluated by the residue theorem considering the only pole at $v=u-2i$. We obtain

\ba
\mathcal{W}_{hex}^{(f\bar f)}&=&\int _{\m{C}_{S}}
\frac{du}{2\pi} \int _{\mathcal{I}} \frac{dv}{2\pi}\mu _f(u) \mu _{f}(v) e^{-\tau E_f(u) +i\sigma p_f (u)}
 e^{-\tau E_f(v) +i\sigma p_f (v)} \times \\
&\times & \frac{4}{(u-v)^2+4} \ \frac{1}{P_{f\bar f}(u|v)P_{\bar f f}(v|u)} - \nn \\
&-& \int_{\m{C}_{S}}\frac{du}{2\pi}\frac{\mu _f(u) \mu _{f}(u-2i)}{P_{f\bar f}(u|u-2i)P_{\bar f f}(u-2i|u)} e^{-\tau E_f(u) +i\sigma p_f (u)}
 e^{-\tau E_f(u-2i) +i\sigma p_f (u-2i)} \nonumber \, .
\ea
The second addendum (denoted $\mathcal{W}^{(M)}_{hex}$) corresponds to the one meson contribution, which dominates the strong coupling limit, since the first contribution is suppressed by a factor $g^{-2}$ coming from the matrix part\footnote {Since $u$ and $v$ vary on non intersecting domains, after rescaling $u=2g\bar u$, $v=2g\bar v$ the quantity $(u-v)^2$ is always large and scales with $g^2$.} and will contribute to one-loop. The identification becomes even more evident upon introducing the center of mass coordinate $u_M=u-i$ and the mesonic measure
\be
\mu _M(u_M)=-\frac{\mu _f (u_M+i) \mu _f (u_M-i)}{P_{f\bar f}(u_M+i|u_M-i)P_{\bar f f}(u_M-i|u_M+i)} \ ,
\ee
as well as the mesonic energy and momentum
\be
E_M(u_M)=E_f(u_M+i) + E_f(u_M-i) \, , \quad p_M(u_M)=p_f(u_M+i) + p_f(u_M-i) \, .
\ee
In fact, all these notations allow us to rewrite the previous addendum as a genuine one particle contribution
\be\label{Wfbf}
\mathcal{W}^{(M)}_{hex}=\int _{\m{C}_{S}-i}\frac{du_M}{2\pi} \mu _M(u_M) e^{-\tau E_M(u_M) +i\sigma p_M (u_M)} \ ,
\ee
although we shall stress the meson is not, strictly speaking, a proper particle at finite coupling, rather a virtual one, as it does not live in the physical domain. Yet, it acquires the status of actual real particle in the string regime, {\it i.e.} at infinite coupling. Importantly, this procedure confirms the validity -- assumed in \cite{FPR2} -- of the usual $S$-matrix fusion in this peculiar context of the Wilson loop OPE. In details, we can enter the strong coupling regime (so far, everything was for {\it any} coupling $g$), recall that at LO
\be
\mathcal{W}_{hex}^{(f\bar f)}\simeq \mathcal{W}^{(M)}_{hex}
\ee
and neglect the shift $-i$ in the integration contour of (\ref{Wfbf}), as the modulus of the integration variable is bigger that $2g$. Then, we reasonably assume that the leading contribution to (\ref {Wfbf}) will come from the natural (string) domain,which is obtainded by first re-scaling the integration variable $u_M= 2g\coth 2\theta $, then going to the limit $g\rightarrow \infty$. In this regime, at LO the mesonic measure $\mu _M(u) \simeq -1$, and, consequently, (\ref{Wfbf}) becomes
\be\label{1mes}
\mathcal{W}^{(M)}_{hex}\simeq\frac{\sqrt{\lambda}}{2\pi}  \int _{-\infty+i\epsilon}^{+\infty+i\epsilon} \frac{d\theta}{\pi}\,
\frac{e^{-2\tau \cosh \theta +2i\sigma \sinh \theta}}{\sinh ^2 2\theta} \, ,
\ee
which confirms the $m=1$ contribution in the second line of (11.13) in \cite {FPR2}. Note that we may also perform the integration in (\ref{Wfbf}), because of absence of poles, over $\mathcal{I}-i$ (see Figure \ref{CsmallShrinks}) and thus we may write $\mathcal{W}^{(M)}_{hex}$ as an integral over the gluon-like hyperbolic variable $u=2g\tanh 2\theta $. This corresponds to shifting the integration contour of (\ref{1mes}) by $+i\pi/4$ and amounts to

\be
\mathcal{W}^{(M)}_{hex}=-\frac{\sqrt{\lambda}}{2\pi}  \int _{-\infty}^{+\infty} \frac{d\theta}{\pi}\,
\frac{e^{-\sqrt{2}\tau (\cosh \theta+i\sinh \theta) +\sqrt{2}i\sigma (\sinh \theta +i\cosh \theta)}}{\cosh ^2 2\theta} \, .
\ee

\medskip

Now, we pass on to the next case with two small fermions, whose rapidities are called $u_1,\,u_2$, and two small antifermions with rapidities $v_1,\,v_2$, as we wish to obtain a bound state of two mesons. Referring to (\ref{split}), the dynamical factor reads
\ba\label{dyn1}
&& \Pi^{(2)}_{dyn}(u_1,u_2,v_1,v_2) =
\frac{1}{P_{ff}(u_1|u_2)P_{ff}(u_2|u_1)}
\frac{1}{P_{f\bar f}(u_1|v_1)P_{\bar ff}(v_1|u_1)} \times\\
&\times &
\frac{1}{P_{f\bar f}(u_1|v_2)P_{\bar ff}(v_2|u_1)}
\frac{1}{P_{f\bar f}(u_2|v_1)P_{\bar ff}(v_1|u_2)}
\frac{1}{P_{f\bar f}(u_2|v_2)P_{\bar ff}(v_2|u_2)}
\frac{1}{P_{\bar f\bar f}(v_1|v_2)P_{\bar f\bar f}(v_2|v_1)} \ , \nn
\ea
while the matrix factor takes the form
\ba\label{Pi2mat}
&& \Pi_{mat}^{(2)}(u_{1},u_{2},v_{1},v_{2}) =\frac{1}{8}\int_{-\infty}^{+\infty} \frac{da_1 db_1 dc_1 da_2 db_2 dc_2}{(2\pi)^6}
\,\frac{g(a_1-a_2) g(b_1-b_2) g(c_1-c_2)}{f(a_1-b_1) f(a_1-b_2) f(a_2-b_1) f(a_2-b_2)} \nn\\
&& \times \frac{1}{f(c_1-b_1) f(c_1-b_2) f(c_2-b_1) f(c_2-b_2)}
\frac{1}{f(u_1-a_1) f(u_1-a_2) f(u_2-a_1) f(u_2-a_2)} \times \nn\\
&& \times \frac{1}{f(v_1-c_1) f(v_1-c_2) f(v_2-c_1) f(v_2-c_2)} \ .
\ea
$\Pi_{mat}^{(2)}$ in (\ref{Pi2mat}) can be worked out by residue method so that it can be given the following structure ({\it cf.} also the independent work \cite{BEL} with the \textit{caveat} that we are here considering the same number of fermions and antifermions):
\ba\label{Pi_mat0}
& \Pi_{mat}^{(2)}(u_{1},u_{2},v_{1},v_{2}) = 4\,\frac{24+3[(u_1-v_1)^2+6][(u_2-v_2)^2+6]+3[(u_1-v_2)^2+6][(u_2-v_1)^2+6]+[(u_1-u_2)^2+4][(v_1-v_2)^2+4]}{[(u_1-u_2)^2+1][(v_1-v_2)^2+1][(u_1-v_1)^2+4][(u_1-v_2)^2+4][(u_2-v_1)^2+4][(u_2-v_2)^2+4]} \, .
\nn\\
\ea
Once (\ref{Pi_mat0}) is plugged into the four fermion contribution to the hexagonal Wilson loop (\ref{WLferm}), the latter can be compactly recast into the shape
\be\label{Whexffff}
\mathcal{W}^{(ff\bar{f}\bar{f})}_{hex}=\frac{1}{4}\int _{\m{C}_{S}}\frac{du_1du_2dv_1dv_2}{(2\pi)^4}
\Phi(u_1,u_2,v_1,v_2)\Pi_{mat}^{(2)}(u_1,u_2,v_1,v_2) \ .
\ee
The function $\Phi$ is given by
\ba
\Phi(u_1,u_2,v_1,v_2) &\equiv & \frac{\hat{\mu}_f(u_1)\hat{\mu}_f(u_2)\hat{\mu}_{\bar{f}}(v_1)\hat{\mu}_{\bar{f}}(v_2)}{P_{ff}(u_1|u_2)P_{ff}(u_2|u_1)P_{ff}(v_1|v_2)P_{ff}(v_2|v_1)P_{f\bar f}(u_1|v_1)P_{f\bar f}(v_1|u_1)}\times \nn\\
&\times & \frac{1}{P_{f\bar f}(u_1|v_2)P_{f\bar f}(v_2|u_1)P_{f\bar f}(u_2|v_1)P_{f\bar f}(v_1|u_2)P_{f\bar f}(u_2|v_2)P_{f\bar f}(v_2|u_2)} \, ,  \label {Phi-def}
\ea
where we used the shorthand notation $ \hat \mu _f(u) = \mu _f(u) e^{-\tau E_f(u)+i\sigma p_f(u)}$.
It is an important remark that (\ref {Phi-def}) is endowed with double zeroes, due to the presence of a simple pole in the amplitudes $P_{ff}(u|v)$ for coinciding rapidities $v=u$:
\ba\label{zeros}
&\Phi(u,u,v_1,v_2)=\Phi(u_1,u_2,v,v)=0 \ .
\ea

As in the case before, upon adding (and subtracting) integrations of $v_2$ and $v_1$ over $-\cal I$, the leading contribution to
\be
\mathcal{W}^{(ff\bar{f}\bar{f})}_{hex}\simeq \m{W}^{(MM)}_{hex}
\ee
is obtained by performing the integration in $v_2$ and $v_1$ over the closed contour $\m{C}_{HM}$: what is left over is sub-dominant. Hence, the leading part reads
\be
\m{W}^{(MM)}_{hex}=
\frac{1}{2}\int _{\m{C}_{S}}\frac{du_1du_2}{(2\pi)^2}\frac{\Phi(u_1,u_2,u_1-2i,u_2-2i)}{(u_1-u_2)^2[(u_1-u_2)^2+1]} \ .\nn
\ee
The fusion relation (\ref{PMM}) can be set {\it formally} at any coupling and thus entails
\be
\Phi(u+i,v+i,u-i,v-i)=(u-v)^2[(u-v)^2+1]\frac{\hat \mu_M(u)\hat \mu_M(v)}{P^{(MM)}(u|v)P^{(MM)}(v|u)} \, ,
\ee
where $ \hat \mu _M(u) = \mu _M(u) e^{-\tau E_M(u)+i\sigma p_M(u)}$. Therefore, at LO we are able to trade the presence of fermionic quantities for mesonic measures and amplitudes, and hence we are left with the expression
\be\label {4f-fin}
\m{W}^{(MM)}_{hex}=\frac{1}{2}\int _{\m{C}_{S}}\frac{du}{2\pi} \int _{\m{C}_{S}}\frac{dv}{2\pi}\frac{\hat \mu _M(u-i)\hat \mu _M(v-i)}{P^{(MM)}(u-i|v-i)P^{(MM)}(v-i|u-i)} \ .
\ee
Formula (\ref{4f-fin}) is the starting point for our search of bound states of mesons, alternatively to the fusion procedure of $S$-matrices which brought to (\ref{PMM}) in \cite{FPR2}. In fact, the integrand has zeroes and poles, as we can make more manifest if we single out explicitly the simple zero and simple pole of $P^{(MM)}$ (according to Appendix \ref{allcouplings}), by separating them from the regular part $P^{(MM)}_{reg}$  (which instead never vanishes nor diverges),
\be
P^{(MM)}(u|v)=\frac{u-v+i}{u-v} P^{(MM)}_{reg}(u|v) \ .
\ee
This entails
\be
\frac{1}{P^{(MM)}(u|v)P^{(MM)}(v|u)}= \left [1-\frac{1}{(u-v)^2+1} \right ] \frac{1}{P^{(MM)}_{reg}(u|v)P^{(MM)}_{reg}(v|u)} \ .
\label{splitting}
\ee
As a result, (\ref {4f-fin}) can be split into a sum of two parts
\be
\m{W}^{(MM)}_{reg}=\m{W}^{(MM)}_{reg}+\m{W}^{(MM)}_{sing} \, ,
\ee
where in the first one
\ba\label{mmreg}
\m{W}^{(MM)}_{reg}&=&\frac{1}{2}\int _{\m{C}_{S}}\frac{du}{2\pi} \int _{\m{C}_{S}}\frac{dv}{2\pi}\frac{\hat \mu _M(u-i)\hat \mu _M(v-i)}{P^{(MM)}_{reg}(u-i|v-i)P^{(MM)}_{reg}(v-i|u-i)} = \, \label {4f-fin-reg} \nn\\
&=& \frac{1}{2}\int _{\m{C}_{S}-i}\frac{du_M}{2\pi} \int _{\m{C}_{S}-i}\frac{dv_M}{2\pi}\frac{\hat \mu _M(u_M)\hat \mu _M(v_M)}{P^{(MM)}_{reg}(u_M|v_M)P^{(MM)}_{reg}(v_M|u_M)} \ ,
\ea
the integrand is free of poles, so that it can be interpreted as the contribution of two free (\textit{i.e.} not bound) mesons, whilst the second part
\be
\m{W}^{(MM)}_{sing}=-\frac{1}{2}\int _{\m{C}_{S}}\frac{du}{2\pi} \int _{\m{C}_{S}}\frac{dv}{2\pi}\frac{\hat \mu _M(u-i)\hat \mu _M(v-i)}{P^{(MM)}_{reg}(u-i|v-i)P^{(MM)}_{reg}(v-i|u-i)} \frac{1}{(u-v)^2+1} \ , \label {4f-fin-sing}
\ee
exhibits the presence of two simple poles in the integrand for $v=u\pm i$\,. Now, we notice the similarity of this formula with (\ref{fer-antifer}) and can employ the same procedure to obtain $\m{W}^{(MM)}_{sing}$ (\ref {4f-fin-sing}) at LO by evaluating the residue in $v=u-i$:
\be
\m{W}^{(2M)}=\frac{1}{4} \int _{\m{C}_{S}}\frac{du}{2\pi}
\frac{\hat \mu _M(u-i)\hat \mu _M(u-2i)}{P^{(MM)}_{reg}(u-i|u-2i)P^{(MM)}_{reg}(u-2i|u-i)} \label {4f-fin-sing-2} \, .
\ee
A simple redefinition of the integration variable $u_{2M}=u-\frac{3i}{2}$ suggests us to give (\ref{4f-fin-sing-2}) a natural interpretation as a composite state of two mesons, characterised by energy and momentum defined respectively by
\be
E_{2M}(u_{2M})=E_M(u_{2M}+i/2)  + E_{M}(u_{2M}-i/2) \, , \quad p_{2M}(u_{2M})=p_{M}(u_{2M}+i/2) + p_{M}(u_{2M}-i/2) \ ,
\ee
and, then, we are induced to define the measure of this particle in the following way:
\be
\mu _{2M}(u_{2M})=\mu _{2M}\left (u-\frac{3i}{2}\right )=\frac{1}{4} \frac{\mu _M(u-i)\mu _M(u-2i)}{P^{(MM)}_{reg}(u-i|u-2i)P^{(MM)}_{reg}(u-2i|u-i)} \ .
\ee
Eventually, it is straightforward to rewrite (\ref {4f-fin-sing-2}) in the single particle contribution form
\be\label{2m}
\m{W}^{(2M)}= \int _{\m{C}_{S}-3i/2}\frac{du_{2M}}{2\pi} \mu _{2M}(u_{2M})
e^{-\tau E_{2M}(u_{2M}) +i\sigma p_{2M}(u_{2M})} \,.
\ee
Again, we must stress how, as in the case of the meson, this composite particle shall not be intended as a genuine one at finite value of the coupling\footnote{Nevertheless, energy, momentum and the complete set of scattering phases associated to mesons and their bound states can be formally formulated even at finite coupling, as in \cite{FPR2}.}, but at infinite $g$. In this limit, the particle becomes real and contributes by passing to the scaled variable $u=2g\coth 2\theta $, as we argued in discussing the one meson case. In fact, we find
\be
P^{(MM)}_{reg}(\theta |\theta ') P^{(MM)}_{reg} (\theta ' |\theta)= \exp \left [ - \frac{2\pi}{\sqrt{\lambda}}\frac{\sinh 2\theta \sinh 2 \theta '}{\cosh (\theta -\theta ')} \right ] \, ,
\ee
and, putting all pieces together, we obtain
\ba
\m{W}^{(2M)} &=& - \frac{\sqrt{\lambda}}{2\pi} \int _{-\infty+i\epsilon}^{+\infty+i\epsilon} \frac{d\theta}{4\pi \sinh ^2 2\theta}
e^{-4\tau \cosh \theta +4i\sigma \sinh \theta} \label {2mes}\\
\m{W}^{(MM)}_{reg} &=& \frac{1}{2} \left (\frac{\sqrt{\lambda}}{2\pi} \right )^2 \int  _{-\infty+i\epsilon}^{+\infty+i\epsilon} \frac{d\theta}{\pi \sinh ^2 2\theta} \int  _{-\infty+i\epsilon}^{+\infty+i\epsilon} \frac{d\theta '}{\pi \sinh ^2 2\theta '} \left [1+\frac{2\pi}{\sqrt{\lambda}} \frac{\sinh 2\theta \sinh 2\theta '}{\cosh (\theta -\theta ')} \right ] \times \nn\\
&\times & e ^{-2\tau (\cosh \theta + \cosh \theta ') +2i\sigma (\sinh \theta + \sinh \theta ')} \ , \label {mm}
\ea
in agreement with the corresponding formul{\ae} of \cite {FPR2}: (\ref {2mes}) matches the $m=2$ term of the sum in the second line of (11.13), (\ref {mm}) the term $m_1=m_2=1$ of formula below (11.23).

Note that in both formul{\ae} (\ref{mmreg}) and (\ref{2m}), since the integrands have no poles in the small fermion sheet (see the discussion in Appendix \ref{allcouplings}), we can shrink the integration contour to $\mathcal{I}$ in place of $\m{C}_{S}$, so that $\mathcal{W}^{(2M)}$ and $\m{W}^{(MM)}_{reg}$ can be rewritten as integrals over gluon-like hyperbolic variables
\ba
\mathcal{W}^{(2M)}&=&\frac{\sqrt{\lambda}}{2\pi}  \int _{-\infty}^{+\infty} \frac{d\theta}{4\pi\cosh ^2 2\theta}\,
e^{-2\sqrt{2}\tau (\cosh \theta+i\sinh \theta) +2\sqrt{2}i\sigma (\sinh \theta +i\cosh \theta)}  \\
\m{W}^{(MM)}_{reg} &=& \frac{1}{2} \left (\frac{\sqrt{\lambda}}{2\pi} \right )^2 \int _{-\infty}^{+\infty} \frac{d\theta}{\pi \cosh ^2 2\theta} \int _{-\infty}^{+\infty} \frac{d\theta '}{\pi \cosh ^2 2\theta '} \left [1-\frac{2\pi}{\sqrt{\lambda}} \frac{\cosh 2\theta \cosh 2\theta '}{\cosh (\theta -\theta ')} \right ] \times \nn\\
&\times & e ^{-\sqrt{2}\tau (\cosh \theta + i\sinh \theta + \cosh \theta ' +i\sinh \theta ') +\sqrt{2}i\sigma (\sinh \theta + i\cosh \theta + \sinh \theta ' + i\cosh \theta ')} \, .
\ea

\medskip

Some comments and comparisons with earlier paper \cite{FPR2} are now due. The above splitting $\m{W}^{(MM)}_{hex}=\m{W}^{(MM)}_{reg}+
\m{W}^{(MM)}_{sing}$ is very natural if performed at generic coupling.
It becomes trickier when we have to deal with the strong coupling regime. In fact, a safe path to follow is first to integrate over the pole in (\ref {4f-fin-sing}) and then to go to the strong coupling limit as last step. This was illustrated at the end of this section and the final answer -- formul{\ae} (\ref {2mes}, \ref {mm}) -- correctly contains two terms, one coming from the singular, the other from the regular part. In \cite {FPR2} on the other hand, the strong coupling limit was performed at an earlier stage -- at the level of formul{\ae} (\ref {4f-fin}, \ref {splitting}) -- hence the singular term is apparently negligible and one is left with the regular part only. Actually, the contribution from the missing singular part is recovered by directly introducing bound states of mesons at strong coupling (by $S$-matrix fusion or bootstrap): in that picture, the term (\ref {2mes}) appears as coming from a bound state of two mesons. It is worth pointing out the structural difference between the fermion-antifermion (\ref{fer-antifer}) and the meson-meson (\ref{4f-fin}) parts: the former has only the singular part which at the leading order, without the regular one, do not yield an unbound fermion contribution. In other words, the fermions show a sort of {\it confinement} mechanism at infinite coupling.

\medskip


The whole two meson term (\ref{4f-fin}), once taken into account the splitting (\ref{splitting}) 
to highlight the singular and the regular parts of the integrand, closely resembles the contribution coming from two particles (the instantons) in the Nekrasov partition function in some $\mathcal{N} = 2$ theories with an $\Omega$-background parametrised by $\epsilon_1$ and $\epsilon_2$ \cite{NS}. The entire partition function is a sum over the number, $N$, of instantons, each interacting with the others and an external potential. In the so-called Nekrasov-Shatashvili (NS) limit $\epsilon_2 \to 0$, which corresponds to our strong coupling limit, the leading contribution was worked out extensively by \cite{Meneghelli, Bourgine2014} and resulted in a sum over instantons and their bound states so to give rise to a TBA-like equation. In a nutshell, the limiting series shares the spirit of the meson sector with the hexagonal Wilson loop in \cite{FPR2}. In fact, the {\it short-range} interaction part \cite{Bourgine2015b} of the partition function shares exactly the very same shape with the polar part in (\ref{4f-fin-sing}), provided $\epsilon_2\sim i g^{-1}$: as a consequence it originates the two-instanton bound state once we integrate by residues and then send $\epsilon_2$ to zero\footnote{The inverse order would have made this part subleading as it was in \cite{FPR2}.}. Despite the formation of the bound states of two particles (two instantons on one hand, or two mesons in our present case) follows the same pattern, there is a fundamental difference between the two cases: in the NS one the integration contour is closed; in the present case, instead, the contour is open, and, becomes closed upon introducing the additional curve $\m{I}$ (see Figure \ref{Cfigura2}); the latter brings a subleading contribution to the integral\footnote{In the NS partition function, on the other hand, additional subleading terms are generated by the presence of poles in the potential, see {\it Conclusions and outlook}.}. In connection with this issue, another main difference with respect to the Nekrasov partition function is that the meson is a composite object appearing as a bound state of more fundamental particles in the strong coupling limit and, if we are interested in the subleading corrections, we shall consider also the effect of the unbound fermions (which do confine differently from the instantons or mesons).


\medskip

In this section we have brought additional evidence to the claim that at infinite coupling mesons and bound states thereof shall be taken into account in the BSV series. We have found that the terms in the fermion/antifermion sector when evaluated at strong coupling acquire the form of contributions coming from a mass two excitation and its bound states.
The results just achieved thus confirm the assumptions made in \cite {FPR2}, on the basis of which we re-summed at strong coupling the OPE series for the hexagon, finding agreement with the classical string results (TBA). It is natural to complete this work by re-summing the OPE series for a general polygon: a successful comparison with TBA \cite{TBuA,YSA,Anope} will be a further confirmation of the picture of the meson \cite{BSV3,FPR2} and its bound states presented in \cite{FPR2}. This will be done in the following: in next Section we review TBA results and in Section \ref {predictions} we sum the OPE series for a polygon at strong coupling. Comparisons between the two approaches are discussed in Section \ref {comparison}.

\section{Revisiting the classical string results}\label{class-str}
\setcounter{equation}{0}

The main aim of this section is to recast the TBA-like integral equations and the critical Yang-Yang functional, which represent the results of the classical string minimisation and may be found in \cite{YSA, Anope}, into a more suitable shape (formul{\ae} (\ref {tbaeps1}, \ref {tbaeps2}, \ref {tbaeps3}) and (\ref{YYc2})) for comparison with the strong coupling re-summation of the OPE series of next Section \ref{predictions}.

\subsection{TBA equations}

Our study moves its first step from equations (47) of \cite{YSA} (or equivalently equations (F.1) of \cite{Anope}). For our present purpose, we are compelled to write explicitly the integration paths (missing in aforementioned paper formul{\ae}), which are different for each term appearing in the right hand sides and coincide with straight lines parallel to the real axis, \textit{i.e.} $\mathbb{R}+i\varphi_s$; hence the TBA equations read, for clarity:
\ba
\ln Y_{2,s}(\theta)&=& - |m_s| \sqrt{2} \cosh (\theta - i \varphi _s) -\int_{\textrm{Im} \theta ' =\varphi _{s}} d\theta'\biggl[K_2(\theta-\theta')
{\cal L}_{s}(\theta ') + \nonumber \\
&+& 2K_1(\theta-\theta')\tilde {\cal L}_{s}(\theta')\biggr] + \int_{\textrm{Im} \theta ' =\varphi _{s-1}} d\theta'\biggl[K_2(\theta-\theta')
\tilde {\cal L}_{s-1}(\theta') + \nonumber \\
&+& K_1(\theta-\theta'){\cal L}_{s-1}(\theta')\biggr] + \int_{\textrm{Im} \theta ' =\varphi _{s+1}} d\theta'\biggl[K_2(\theta-\theta')
\tilde {\cal L}_{s+1}(\theta') + \nonumber \\
&+& K_1(\theta-\theta'){\cal L}_{s-1}(\theta')\biggr] \, , \label {eq1} \\
\ln Y_{1,s}(\theta)&=& - |m_s| \cosh (\theta - i \varphi _s) -C_s -\int_{\textrm{Im} \theta ' =\varphi _{s}} d\theta'\biggl[K_2(\theta-\theta')
\tilde {\cal L}_{s}(\theta') + \nonumber \\
&+& K_1(\theta-\theta'){\cal L}_{s}(\theta')\biggr] + \int_{\textrm{Im} \theta ' =\varphi _{s-1}}
d\theta'\biggl[K_1(\theta-\theta')\tilde {\cal L}_{s-1}(\theta') + \nonumber \\
&+& \frac{1}{2}K_2(\theta-\theta'){\cal L}_{s-1}(\theta') - \frac{1}{2}K_3(\theta -\theta'){\cal M}_{s-1}(\theta')\biggr] + \nonumber \\
&+&  \int_{\textrm{Im} \theta ' =\varphi _{s+1}} d\theta'\biggl[K_1(\theta-\theta')\tilde {\cal L}_{s+1}(\theta') + \frac{1}{2}K_2(\theta-\theta'){\cal L}_{s+1}(\theta') + \nonumber \\
&+& \frac{1}{2}K_3(\theta -\theta'){\cal M}_{s+1}(\theta')\biggr] \, , \label {eq2} \\
\ln Y_{3,s}(\theta)&=& - |m_s| \cosh (\theta - i \varphi _s) +C_s -\int_{\textrm{Im} \theta ' =\varphi _{s}} d\theta'\biggl[K_2(\theta-\theta')
\tilde {\cal L}_{s}(\theta') + \nonumber \\
&+& K_1(\theta-\theta'){\cal L}_{s}(\theta')\biggr] + \int_{\textrm{Im} \theta ' =\varphi _{s-1}}
d\theta'\biggl[K_1(\theta-\theta')\tilde {\cal L}_{s-1}(\theta') + \nonumber \\
&+& \frac{1}{2}K_2(\theta-\theta'){\cal L}_{s-1}(\theta') + \frac{1}{2}K_3(\theta -\theta'){\cal M}_{s-1}(\theta')\biggr] + \nonumber \\
&+&  \int_{\textrm{Im} \theta ' =\varphi _{s+1}} d\theta'\biggl[K_1(\theta-\theta')\tilde {\cal L}_{s+1}(\theta') + \frac{1}{2}K_2(\theta-\theta'){\cal L}_{s+1}(\theta') - \nonumber \\
&-& \frac{1}{2}K_3(\theta -\theta'){\cal M}_{s+1}(\theta')\biggr] \label {eq3} \, ,
\ea
where, along with the kernels \cite{YSA,Anope}
\be
K_1(\theta )= \frac{1}{2\pi \cosh \theta} \, , \quad K_2(\theta )= \frac{\sqrt{2}\cosh \theta}{\pi \cosh 2\theta}
\, , \quad K_3(\theta )= \frac{i}{\pi} \tanh 2\theta  \label {ker} \, ,
\ee
the following short notations for nonlinear functions of $\ln Y$'s were introduced:
\begin{equation}
 {\cal L}_{s}(\theta)=\ln(1+Y_{1,s}(\theta))(1+Y_{3,s}(\theta)) \, , \quad \tilde {\cal L}_{s}(\theta)=\ln(1+Y_{2,s}(\theta)) \, , \quad {\cal M}_{s}(\theta)=\ln\frac{(1+Y_{1,s}(\theta))}{(1+Y_{3,s}(\theta))} \label {nlinY} \, .
\end{equation}
Equations (\ref {eq1}, \ref {eq2}, \ref {eq3}) depend on the three set of constants $|m_s|, \varphi_s, C_s$, $s=1,...,n-5$, which eventually will be related to the $3n-15$ conformal ratios of a polygonal Wilson loop with $n$ null edges.

The simplest case is offered by the hexagonal Wilson loop, for which the label $s$ sticks to the value $s=1$, so that (\ref {eq1}, \ref {eq2}, \ref {eq3}) coincide with equations (3.6) of \cite{TBuA}, provided one performs the identifications $2Z=|m|$, $\mu =e^{-C}$, in addition to
\be
\epsilon (\theta - i\varphi)=-\ln Y_{1,1}(\theta ) - C \, , \quad \tilde \epsilon (\theta - i\varphi)=-\ln Y_{2,1}(\theta ) \, ,
\ee
together with the relation between the $Y$-functions $\ln Y_{1,1}(\theta )=\ln Y_{3,1}(\theta ) -2C$.

Moving now to the most general case of a polygon with $n$ edges, the label $s$ takes values from $1$ to $n-5$. Let us define the symbol
\be
b_s= \frac{1+(-1)^s}{2} \, , \label {bs}
\ee
i.e. $b_s=1$ for even values of $s$, while $b_s=0$ if $s$ is odd; moreover it turns out convenient to introduce the hatted Y-functions $\hat Y_{\alpha,s} (\theta )$,
\be
\hat Y_{\alpha,s}(\theta )= Y_{\alpha,s} \left ( \theta - \frac{i\pi}{4}b_{\alpha + s + 1} \right )
\label {hatY} \, ,
\ee
which can be put directly into relation with the physical cross-ratios when evaluated at $\q=0$
\be
y_{\alpha,s}= \hat Y_{\alpha,s}(0) \label {cr-rat} \, ,
\ee
and also to define the tilded kernels \cite{Anope}:
\ba
&& \tilde K_1(\theta , \theta ')= - \frac{1}{2\pi} \frac{\sinh 2\theta}{\sinh 2\theta ' \cosh (\theta -\theta ')}
\, , \quad \tilde K_3(\theta , \theta ')=\frac{i}{\pi} \frac{\sinh 2\theta}{\sinh 2\theta ' \sinh (2\theta -2\theta ')} \nonumber \\
&& \tilde K_2^{(s)} (\theta , \theta ')=- \frac{\sqrt{2}}{\pi}\sinh \left (\theta -\theta ' + \frac{i\pi}{4}(-1)^s \right )
\frac{\sinh 2\theta}{\sinh 2\theta ' \sinh (2\theta -2\theta ')}  \, . \label {tildeK}
\ea
In doing so, from (\ref {eq1}, \ref {eq2}, \ref {eq3}) we may obtain the integral equations for the hatted $Y$ functions, which read
\ba\label{hatTBA1}
&& \ln \hat Y_{2,s} (\theta )-{\cal E}_s(\theta)=-\int_{\textrm{Im} \theta ' =\varphi _{s}} d\theta' \biggl[\tilde K_2^{(s)}\left(\theta, \theta' + \frac{i\pi}{4}b_s\right){\cal L}_s (\theta') + \nonumber \\
&+& 2\tilde K_1\left(\theta,\theta' + \frac{i\pi}{4}b_{s+1}\right)\tilde {\cal L}_s(\theta')\biggr] +  \int_{\textrm{Im} \theta ' =\varphi _{s-1}} d\theta ' \biggl[\tilde K_1\left(\theta, \theta '+ \frac{i\pi}{4}b_{s+1}\right) {\cal L}_{s-1}(\theta ') + \nonumber \\
&+& \tilde K_2^{(s)}\left(\theta, \theta' + \frac{i\pi}{4}b_s\right) \tilde {\cal L}_{s-1}(\theta')\biggr]  + \int_{\textrm{Im} \theta ' =\varphi _{s+1}} d\theta ' \biggl[\tilde K_1\left(\theta, \theta '+ \frac{i\pi}{4}b_{s+1}\right) {\cal L}_{s+1}(\theta ') + \nonumber \\
&+& \tilde K_2^{(s)}\left(\theta, \theta' + \frac{i\pi}{4}b_s\right) \tilde {\cal L}_{s+1}(\theta')\biggr] \, ,
\ea
\ba
&& \ln \hat Y_{1,s} (\theta )+ \ln \hat Y_{3,s} (\theta )-\sqrt{2}{\cal E}_s\left (\theta + \frac{i\pi}{4}(-1)^{s+1} \right ) = -\int_{\textrm{Im} \theta ' =\varphi _{s}} d\theta' \biggl[2\tilde K_2^{(s)}\left(\theta, \theta' + \frac{i\pi}{4}b_{s+1}\right)\tilde {\cal L}_s (\theta') + \nonumber \\
&+& 2\tilde K_1\left(\theta,\theta' + \frac{i\pi}{4}b_s\right){\cal L}_s(\theta')\biggr] +  \int_{\textrm{Im} \theta ' =\varphi _{s-1}} d\theta ' \biggl[\tilde K^{(s)}_2\left(\theta, \theta '+ \frac{i\pi}{4}b_{s+1}\right) {\cal L}_{s-1}(\theta ') + \nonumber \\
&+& 2\tilde K_1\left(\theta, \theta' + \frac{i\pi}{4}b_s\right) \tilde {\cal L}_{s-1}(\theta')\biggr] + \int_{\textrm{Im} \theta ' =\varphi _{s+1}} d\theta ' \biggl[\tilde K^{(s)}_2\left(\theta, \theta '+ \frac{i\pi}{4}b_{s+1}\right) {\cal L}_{s+1}(\theta ') + \nonumber \\
&+& 2\tilde K_1\left(\theta, \theta' + \frac{i\pi}{4}b_s\right) \tilde {\cal L}_{s+1}(\theta')\biggr] \, ,  \\
&&  \ln \hat Y_{1,s} (\theta )- \ln \hat Y_{3,s} (\theta ) -\ln y_{1,s} + \ln y_{3,s} =-\int_{\textrm{Im} \theta ' =\varphi _{s-1}} d\theta' \tilde K_3 \left(\theta , \theta'+ \frac{i\pi}{4}b_{s+1}\right) {\cal M}_{s-1}(\theta') + \nonumber \\
&+& \int_{\textrm{Im} \theta ' =\varphi _{s+1}} d\theta' \tilde K_3 \left(\theta , \theta'+ \frac{i\pi}{4}b_{s+1}\right) {\cal M}_{s+1}(\theta')
\label{hatTBA2} \, ,
\ea
where the function ${\cal E}_s(\theta )$ is given by
\ba
{\cal E}_s(\theta )&=& -i (-1)^s \left [\sqrt{2}\sinh \left ( \theta + \frac{i\pi}{4}(-1)^s \right ) \ln y_{2,s}-\sinh \theta \ln y_{1,s}y_{3,s} \right ]  = \label  {Estheta} \, \\
&=& \cosh \theta \ln y_{2,s} + i (-1)^{s+1}\sinh \theta \ln \frac{y_{2,s}}{y_{1,s}y_{3,s}} \, . \nonumber
\ea
Keeping in mind the hexagon case, we define the pseudo-energies $\epsilon_{a,s}$, upon relating them to the $Y$-functions according to
\ba
\epsilon _{1,s}(\theta -i \varphi _s)&=&-\ln Y_{1,s}(\theta) - \frac{1}{2}\ln \frac{y_{3,s}}{y_{1,s}} \, ,  \\
\epsilon _{3,s}(\theta -i \varphi _s)&=&-\ln Y_{3,s}(\theta) + \frac{1}{2}\ln \frac{y_{3,s}}{y_{1,s}} \, , \\
\epsilon _{2,s}(\theta -i \varphi _s)&=&-\ln Y_{2,s}(\theta) \, .
\ea
although it will turn out useful to express the $\epsilon_{a,s}$ functions also in terms of the hatted-Y
\ba
\epsilon _{1,s}(\theta -i \hat \varphi _s)&=&-\ln \hat Y_{1,s}\left (\theta - \frac{i\pi}{4}b_{s+1} \right ) - \frac{1}{2}\ln \frac{y_{3,s}}{y_{1,s}} \, , \\
\epsilon _{3,s}(\theta -i \hat \varphi _s)&=&-\ln \hat Y_{3,s}\left (\theta - \frac{i\pi}{4}b_{s+1} \right ) + \frac{1}{2}\ln \frac{y_{3,s}}{y_{1,s}} \, , \\
\epsilon _{2,s}(\theta -i \hat \varphi _s)&=&-\ln \hat Y_{2,s}\left (\theta - \frac{i\pi}{4}b_s \right ) \, ,
\ea
where we defined the quantity
\be
\hat \varphi _s= \varphi _s + \frac{\pi}{4} \, .
\ee
Finally, we introduce the functions
\be
L_s(\theta )=\ln \left [\left ( 1+\sqrt{\frac{y_{1,s}}{y_{3,s}}}e^{-\epsilon _{1,s}(\theta -i\hat \varphi _s)}\right )\left (1+\sqrt{\frac{y_{3,s}}{y_{1,s}}}e^{-\epsilon _{3,s}(\theta -i\hat \varphi _s)}\right ) \right ]={\cal L}_s\left(\theta -\frac{i\pi}{4}\right) \, ,
\ee
\be
\tilde L_s(\theta )= \ln \left (1+e^{-\epsilon _{2,s}(\theta -i\hat \varphi _{s})}\right )=\tilde {\cal L}_s\left(\theta -\frac{i\pi}{4}\right) \, ,
\ee
\be
M_s(\theta )=\ln \left ( \frac{1+\sqrt{\frac{y_{1,s}}{y_{3,s}}}e^{-\epsilon _{1,s}(\theta -i\hat \varphi _s)}}{ 1+\sqrt{\frac{y_{3,s}}{y_{1,s}}}e^{-\epsilon _{3,s}(\theta -i\hat \varphi _s)}} \right )={\cal M}_s\left(\theta -\frac{i\pi}{4}\right)
\ee
and rewrite (\ref{hatTBA1}-\ref{hatTBA2}) in terms of the pseudo-energies $\epsilon_{\alpha,s}$:
\small
\ba
&& \epsilon _{2,s}(\theta -i \hat \varphi _s)=-{\cal E}_s \left ( \theta - \frac{i\pi}{4}b_s \right ) + \label{tbaeps1}  \\
&+& \int_{\textrm{Im} \theta ' =\hat\varphi _{s}}d\theta'\biggl[\tilde K_2^{(s)} \left ( \theta - \frac{i\pi}{4}b_s, \theta '- \frac{i\pi}{4}b_{s+1} \right ) L_s(\theta') + 2\tilde K_1 \left ( \theta - \frac{i\pi}{4}b_s, \theta ' - \frac{i\pi}{4}b_s\right )\tilde L_s(\theta')\biggr] - \nonumber \\
&-& \int_{\textrm{Im} \theta ' =\hat\varphi _{s-1}}d\theta'\biggl[\tilde K_2^{(s)} \left ( \theta - \frac{i\pi}{4}b_s, \theta '- \frac{i\pi}{4}b_{s+1} \right )\tilde L_{s-1}(\theta') + \tilde K_1 \left ( \theta - \frac{i\pi}{4}b_s, \theta ' - \frac{i\pi}{4}b_s\right)L_{s-1}(\theta')\biggr] - \nonumber \\
&-& \int_{\textrm{Im} \theta ' =\hat\varphi _{s+1}}d\theta'\biggl[\tilde K_2^{(s)} \left ( \theta - \frac{i\pi}{4}b_s, \theta '- \frac{i\pi}{4}b_{s+1} \right )\tilde L_{s+1}(\theta') + \tilde K_1 \left ( \theta - \frac{i\pi}{4}b_s, \theta ' - \frac{i\pi}{4}b_s\right)L_{s+1}(\theta')\biggr] \nonumber \\
&& \epsilon _{3,s}(\theta -i \hat \varphi _s)-\epsilon _{1,s}(\theta -i \hat \varphi _s)= -\int_{\textrm{Im} \theta ' =\hat\varphi _{s-1}}d\theta'\biggl[ \tilde K_3 \left ( \theta - \frac{i\pi}{4}b_{s+1}, \theta ' - \frac{i\pi}{4}b_s\right ) M_{s-1}(\theta')\biggr] + \label{tbaeps2}  \\
&+& \int_{\textrm{Im} \theta ' =\hat\varphi _{s+1}}d\theta'\biggl[ \tilde K_3 \left ( \theta - \frac{i\pi}{4}b_{s+1}, \theta ' - \frac{i\pi}{4}b_s\right ) M_{s+1}(\theta')\biggr] \nonumber \\
&& \epsilon _{3,s}(\theta -i \hat \varphi _s)+\epsilon _{1,s}(\theta -i \hat \varphi _s)= -\sqrt{2} {\cal E}_s \left ( \theta - \frac{i\pi}{4}b_s \right ) + \label{tbaeps3} \\
&+& 2\int_{\textrm{Im} \theta ' =\hat\varphi _{s}}d\theta'\biggl[ \tilde K_1 \left ( \theta - \frac{i\pi}{4}b_{s+1}, \theta ' - \frac{i\pi}{4}b_{s+1}\right )L_s(\theta') +
 \tilde K_2^{(s)} \left ( \theta - \frac{i\pi}{4}(-1)^s-\frac{i\pi}{4}b_s, \theta ' - \frac{i\pi}{4}b_s \right )\tilde L_s(\theta')\biggr] - \nonumber \\
&-&  \int_{\textrm{Im} \theta ' =\hat\varphi _{s-1}}d\theta'\biggl[2 \tilde K_1 \left ( \theta - \frac{i\pi}{4}b_{s+1}, \theta ' - \frac{i\pi}{4}b_{s+1}\right )\tilde L_{s-1}(\theta') + \tilde K_2^{(s)} \left ( \theta - \frac{i\pi}{4}(-1)^s -\frac{i\pi}{4}b_s, \theta ' - \frac{i\pi}{4}b_s \right ) L_{s-1}(\theta')\biggr] - \nonumber \\
&-&  \int_{\textrm{Im} \theta ' =\hat\varphi _{s+1}}d\theta'\biggl[2 \tilde K_1 \left ( \theta - \frac{i\pi}{4}b_{s+1}, \theta ' - \frac{i\pi}{4}b_{s+1}\right )\tilde L_{s+1}(\theta') + \tilde K_2^{(s)} \left ( \theta - \frac{i\pi}{4}(-1)^s -\frac{i\pi}{4}b_s, \theta ' - \frac{i\pi}{4}b_s \right ) L_{s+1}(\theta')\biggr] \nonumber \, .
\ea
\normalsize

\subsection{Yang-Yang functional}

We now turn our attention to the Yang-Yang functional: following \cite{Anope}, the equations (\ref {tbaeps1}, \ref {tbaeps2}, \ref {tbaeps3}) can be compactly rewritten by introducing the functions $\hat A_{\alpha,s}(\theta)$ according to:
\ba
&& \epsilon _{2,s}(\theta -i \hat \varphi _s) + {\cal E}_s \left ( \theta - \frac{i\pi}{4}b_s \right )
= -\hat A_{2,s}(\theta) \, ,
\ea
\ba
&& \epsilon _{3,s}(\theta -i \hat \varphi _s)-\epsilon _{1,s}(\theta -i \hat \varphi _s)=
\hat A_{1,s}(\theta) - \hat A_{3,s}(\theta) \, ,
\ea
\ba
&& \epsilon _{3,s}(\theta -i \hat \varphi _s)+\epsilon _{1,s}(\theta -i \hat \varphi _s)
 + \sqrt{2} {\cal E}_s \left ( \theta - \frac{i\pi}{4}b_s \right )
 =  -\hat A_{3,s}(\theta) - \hat A_{1,s}(\theta) \, .
\ea
In doing so, we are able to reformulate the extremal value of the Yang-Yang functional, whose
expression\footnote{We remark that, with respect to formula (F.8) of \cite{Anope} in (\ref {YYc2}) we changed the sign of the second term in the square bracket. We believe that there is a typographic error in (F.8), since subsequent relations (F.42-F.46) are compatible with a plus sign in front of the second term in the square bracket.}
was originally given by \cite{Anope}, in terms of the functions $\epsilon _{a,s}(\theta -i\hat \varphi _s)$ and $\hat A_{a,s}(\theta)$:
\ba\label{YYc2}
YY_c&=&\sum _{\alpha=1}^3 \sum _{s=1}^{n-5} \int _{\textrm{Im} \theta =\hat \varphi _s}\frac{d\theta}{\pi \sinh ^2 \left [2\theta-
\frac{i\pi}{2}b_{\alpha + s}\right ] }  \Bigl [ \textrm{Li}_2 \left (-e^{-\epsilon _{\alpha,s}(\theta -i\hat \varphi _s)} \left ( \frac{y_{1,s}}{y_{3,s}} \right )^{1-\frac{\alpha}{2}} \right )+ \nonumber \\
&+&  \frac{1}{2}\ln \left ( 1+e^{-\epsilon _{\alpha,s}(\theta -i\hat \varphi _s)} \left ( \frac{y_{1,s}}{y_{3,s}}\right )^{1-\frac{\alpha}{2}}  \right ) \hat A_{\alpha,s} (\theta ) \Bigr ]  \, .
\ea
It is worth reminding that the critical value of the Yang-Yang potential is related to the strong coupling limit of the conformally invariant finite ratio for null polygonal Wilson loops ${\cal W}_n$ \cite{BSV1} according to the relation
\be
{\cal W}_n =e^{-\frac{\sqrt{\lambda}}{2\pi} YY_c} \, . \label {W-YY}
\ee
The aim of next two sections is to verify statement (\ref {W-YY}) for a general polygon. We start in next section by
computing the left hand side of (\ref {W-YY}) through the re-summation of the OPE series of Basso, Sever and Vieira at strong coupling.

\section{Re-summing the OPE series with mesons (and bound states)}\label{predictions}
\setcounter{equation}{0}

As previously announced, we are now going to show how the OPE spectral series written in terms of pentagon transitions \cite{BSV1}-\cite{BSV5}, for an arbitrary null polygonal Wilson loops, can be re-summed at strong coupling ($\lambda\longrightarrow\infty$) to the (exponential of) Yang-Yang functional given in \cite{Anope}, determined through the solution of the TBA integral equations \cite{TBuA}-\cite{Anope}. We regard as our pivotal hypothesis that the relevant excitations to be taken into account are gluons and their bound states together with mesons (and bound states thereof), which should be meant, as exhaustively argued in Section \ref{birth}, as particles corresponding to bound states of small fermions and antifermions. In this section and in the following we thus recall and generalise what was done in \cite{FPR2}.
In the first place, we wish to recollect some useful results. An explanation on the notation adopted is due: the vectorial notation $\vec{A}=(A_1,...,A_N)$, where $A_i=a_{\alpha _i}$,
indicates that the $i$-th excitation is a bound state of $a_{\alpha_i}$ particles of type $\alpha_i$, while the index $\alpha_i$ takes the values $\alpha_i=1,\,3$ when referred to gluons of positive and negative helicity, respectively, while $\alpha _i=2$ denotes mesons.

The pentagonal amplitude between an intermediate state of $N$ excitations and an intermediate state of $M$ excitations $P_{\vec{A} \vec{B}}(\theta_1,\ldots, \theta_N|\theta'_1,\ldots, \theta'_M)$ can be factorised in terms of transitions between one particle states $P_{A_i B_j}(\theta_i|\theta'_j)$ \cite{BSV1},\cite{BSV5}:
\be
P_{\vec{A} \vec{B}}(\theta_1,\ldots, \theta_N|\theta'_1,\ldots, \theta'_M) = \frac{\prod \limits _{i,j}P_{A_i B_j}(\theta_i|\theta'_j)}{\prod \limits _{i>j}P_{A_i A_j}(\theta_i|\theta_j) \prod \limits _{i<j}P_{B_i B_j}(\theta'_i|\theta'_j)} \, ; \label {P-singl}
\ee
it shall be pointed out that the simple form of the relation above stems from the fact that gluons and mesons (and bound states), the only particles to be considered under the limit $\lambda\longrightarrow\infty$ coupling, behave as singlets under the $SU(4)$ R-symmetry, otherwise a further rational factor would appear. Moreover, when the strong coupling regime is considered, the pentagonal amplitudes of bound states of both gluons or mesons can be formulated in terms of their elementary components \cite{FPR2}, according to the relation
\be
P_{A_iA_j}(\theta|\theta')=[P_{\alpha _i \alpha _j}(\theta|\theta')]^{a_{\alpha _i} a_{\alpha _j}} \, , \label
{P-strong}
\ee
where the 'fundamental' $P_{\alpha  \beta} (\theta | \theta ')$ are listed in Appendix \ref {appA2}. Analogously, energy and momentum of bound states at strong coupling equal the sum of energy and momentum of their constituents, \textit{i.e.}
\be
E_{A_i}(\theta )=a_{\alpha _i}E_{\alpha _i}(\theta) \, , \quad p_{A_i}(\theta )=a_{\alpha _i}p_{\alpha _i}(\theta) \, ,
\ee
reminding the expressions
\be
E_1(\theta)=E_3(\theta)=\sqrt{2}\cosh \theta \, , \quad E_2(\theta)=2\cosh \theta \, ; \quad
p_1(\theta)=p_3(\theta)=\sqrt{2}\sinh \theta \, , \quad p_2(\theta)=2\sinh \theta \label {en-mom} \, .
\ee

\subsection{Heptagon}
We are going to outline with some detail the case of the heptagonal Wilson loop ($n=7$), in order to explain the method employed before turning to the most general polygon. The properties discussed above allow us to write the OPE series for the heptagonal Wilson loop at strong coupling in the form
\ba
{\cal W}_{hep}\equiv{\cal W}_{7}&=&\sum _{N=0}^{\infty} \sum _{M=0}^{\infty}\frac{1}{N!}\frac{1}{M!}\sum_{\alpha_1=1}^3...\sum_{\alpha_N=1}^3 \sum_{\beta_1=1}^3...\sum_{\beta_M=1}^3
\sum _{a_{\alpha _1}}...\sum _{a_{\alpha _N}} \sum _{b_{\beta _1}}...\sum _{b_{\beta _M}} \int \prod _{i=1}^N d\hat \theta_i^{(1)} (\tau _1, \sigma _1, \phi _1) \times \nonumber \\
&\times  & \prod _{j=1}^M d\hat \theta_j^{(2)} (\tau _2, \sigma _2, \phi _2)
\frac{\prod \limits _{i=1}^N \prod \limits _{j=1}^M [P _{\alpha_i \beta_j}(-\theta_i^{(1)}|\theta_j^{(2)})]^{a_{\alpha _i}b_{\beta _j}}}{\prod \limits _{\stackrel {i,j=1}{i\not=j}}^{N} [P_{\alpha_i \alpha_j} (\theta_i^{(1)}|\theta_j^{(1)})]^{a_{\alpha _i}a_{\alpha _j}} \prod \limits _{\stackrel {i,j=1}{i\not=j}}^{M} [P_{\beta_i \beta_j} (\theta_i^{(2)}|\theta_j^{(2)})]^{b_{\beta _i}b_{\beta _j}}} \, , \label {Whep}
\ea
where, for brevity, the measures
\be
\mu _1(\theta)=\mu _3(\theta)= - \frac{\sqrt{\lambda}}{2\pi} \frac{2}{\cosh ^2 2\theta} \, , \quad
\mu _2(\theta)=  \frac{\sqrt{\lambda}}{2\pi} \frac{2}{\sinh ^2 2\theta}
\ee
and the propagation phases have been gathered and merged into the differential, resulting in the short-hand notation
\ba \label {dhat}
d\hat \theta_i^{(1)} (\tau _1, \sigma _1, \phi _1)&=&e^{-\tau _1 a_{\alpha _i}E_{\alpha_i}(\theta_i^{(1)})+i\sigma _1 a_{\alpha _i} p_{\alpha_i}(\theta_i^{(1)})+i a_{\alpha _i} \phi _1(2-\alpha _i)}\frac{\mu _{\alpha _i}(\theta_i^{(1)})}{\left (a_{\alpha _i} \right )^2}(-1)^{a_{\alpha _i}-1} \frac{d\theta_i^{(1)}}{2\pi} \ , \\
d\hat \theta_j^{(2)} (\tau _2, \sigma _2, \phi _2)&=&e^{-\tau _2b_{\beta _j}E_{\beta_j}(\theta_j^{(2)})+i\sigma _2 b_{\beta _j} p_{\beta_j}(\theta_j^{(2)})+i b_{\beta _j} \phi _2 (2-\beta _j)}\frac{\mu _{\beta _j}(\theta_j^{(2)})}{\left (b_{\beta _j} \right )^2}(-1)^{b_{\beta _j}-1} \frac{d\theta_j^{(2)}}{2\pi} \ .  \nn
\ea
We are thus ready to show that the Wilson loop may be rewritten in the shape of a partition function.
In the first place, one can introduce the fields $X_{\alpha}^{(s)}(\theta)$ by means of their propagator $G_{\alpha ,\beta}^{(s,s')}(\theta , \theta ')$
\be
\langle X_{\alpha}^{(s)}(\theta) X_{\beta}^{(s')}(\theta ') \rangle = G_{\alpha ,\beta}^{(s,s')}(\theta , \theta ') \ ,
\label {fi}
\ee
upon fixing its elements
\ba\label {Gchoice}
&& G^{(1,1)}_{\alpha, \beta}(\theta,\theta')=G^{(2,2)}_{\alpha, \beta}(\theta,\theta')=-\ln [P_{\alpha \beta}(\theta|\theta')P_{\beta \alpha}(\theta'|\theta)]\, , \\
&& G^{(1,2)}_{\alpha , \beta}(\theta,\theta')=G^{(2,1)}_{\beta ,\alpha}(\theta',\theta)=-\ln [P_{\beta \alpha}(\theta'|\theta)] \ , \nn
\ea
specifying that the kernels $G^{(s,s')}_{\alpha, \beta}(\theta ,\theta ')$ have support on $\gamma_s \times\gamma _{s'}$, where $\gamma_s$ is a curve defined by the condition $\textrm{Im}\theta=\hat\varphi_{s}-\frac{\pi}{4\,}b_s$. The symbol $\langle \cdot \cdot \cdot \rangle $ in (\ref {fi}) stands for a functional integration with respect to basic fields $X_{\alpha}^{(s)}(\theta)$, which means explicitly
\be
\langle \m{O} \,\rangle=\int  \prod  _{\alpha , \beta =1}^3 {\cal D}X^{(1)}_{\alpha } {\cal D}X^{(2)}_{\beta } \,\m{O}[X^{(1)},X^{(2)}]
\,e^{-S_0[X^{(1)},X^{(2)}]} \, ,
\ee
evaluated by making use of an action involving a quadratic term
\be
S_0[X^{(1)},X^{(2)}] = \frac{1}{2} \sum _{s,s',\alpha ,\beta} \int _{\gamma _{s}} d\theta \int _{\gamma _{s'}} d\theta '  X^{(s)}_{\alpha}(\theta ) T^{(s,s')}_{\alpha, \beta}(\theta , \theta ') X^{(s')}_{\beta}(\theta ') \ , \label {Szero}
\ee
with $T$ the inverse of $G$. For our purposes
it is convenient to introduce the currents $J^{(s)}_{\alpha}(\theta)$, which have support on the straight line $\gamma_s$ and enjoy the form
\be
J^{(1)}_{\alpha}(\theta)=\sum _{i=1}^N a_{\alpha _i} \delta _{\alpha, \alpha _i}\delta (\theta-\theta_i^{(1)})   \, , \quad J^{(2)}_{\beta }(\theta)=-\sum _{j=1}^M b_{\beta _j} \delta _{\beta, \beta _j}\delta (\theta+\theta_j^{(2)}) \label {Jchoice} \ .
\ee
Then, on the one hand, if we work out (\ref{Whep}) using definitions (\ref{Gchoice},\ref{Jchoice}) and property $P_{\alpha \beta}(-\theta|-\theta')=P_{\beta \alpha}(\theta'|\theta)$, we get the following relation for the multi-particle amplitude:
\ba
&& \frac{\prod \limits _{i=1}^N \prod \limits _{j=1}^M [P _{\alpha_i \beta_j}(-\theta_i^{(1)}|\theta_j^{(2)})]^{a_{\alpha _i}b_{\beta _j}}}{\prod \limits _{\stackrel {i,j=1}{i\not=j}}^{N} [P_{\alpha_i \alpha_j} (\theta_i^{(1)}|\theta_j^{(1)})]^{a_{\alpha _i}a_{\alpha _j}} \prod \limits _{\stackrel {i,j=1}{i\not=j}}^{M} [P_{\beta_i \beta_j} (\theta_i^{(2)}|\theta_j^{(2)})]^{b_{\beta _i}b_{\beta _j}}}= \exp \Bigl [\frac{1}{2}\sum _{i,j=1}^N a_{\alpha _i} a_{\alpha _j} G^{(1,1)}_{\alpha _i^{(1)}, \alpha _j^{(1)}}(\theta_i^{(1)},\theta_j^{(1)}) + \\
&& + \frac{1}{2}\sum _{i,j=1}^M b_{\beta _i} b_{\beta _j} G^{(2,2)}_{\beta _i, \beta _j}(-\theta_i^{(2)},-\theta_j^{(2)})- \frac{1}{2}\sum _{i=1}^N \sum _{j=1}^M \Bigl (a_{\alpha _i} b_{\beta _j} G^{(1,2)}_{\alpha _i, \beta _j}(\theta_i^{(1)},-\theta_j^{(2)})+a_{\alpha _i} b_{\beta _j} G^{(2,1)}_{\beta _j, \alpha _i}(-\theta_j^{(2)},\theta_i^{(1)}) \Bigr ) \Bigr ] = \nn\\
&& = \exp \left [ \frac{1}{2}\sum _{s,s', \alpha, \beta}\int _{\gamma _s}d\theta \int _{\gamma _{s'}}d\theta' J^{(s)}_{\alpha}(\theta) G^{(s,s')}_{\alpha, \beta}(\theta,\theta') J^{(s')}_{\beta }(\theta') \right ] \ . \label {iden}
\ea
On the other hand, by coupling currents to fields via linear source terms
\be
\sum _{s,\alpha} \int _{\gamma _s}d\q\, X_{\alpha}^{(s)}(\theta) J_{\alpha}^{(s)}(\theta) \ ,
\ee
the same quantity (\ref {iden}) can be obtained by performing functional (gaussian) integrations
\be
\langle \exp \left [ \sum _{s,\alpha} \int _{\gamma _s}d\q\, X_{\alpha}^{(s)}(\theta) J_{\alpha}^{(s)}(\theta) \right ] \rangle =\exp \left [ \frac{1}{2}\sum _{s,s', \alpha, \beta}\int _{\gamma _s}d\theta \int _{\gamma _{s'}}d\theta' J^{(s)}_{\alpha}(\theta) G^{(s,s')}_{\alpha, \beta}(\theta,\theta') J^{(s')}_{\beta }(\theta') \right ]\label {identity} \, .
\ee
Hence, by means of (\ref {iden}, \ref {identity}) and then (\ref {Jchoice}), we find
\ba
&&  \frac{\prod \limits _{i=1}^N \prod \limits _{j=1}^M [P _{\alpha_i \beta_j}(-\theta_i^{(1)}|\theta_j^{(2)})]^{a_{\alpha _i}b_{\beta _j}}}{\prod \limits _{\stackrel {i,j=1}{i\not=j}}^{N} [P_{\alpha_i \alpha_j} (\theta_i^{(1)}|\theta_j^{(1)})]^{a_{\alpha _i}a_{\alpha _j}} \prod \limits _{\stackrel {i,j=1}{i\not=j}}^{M} [P_{\beta_i \beta_j} (\theta_i^{(2)}|\theta_j^{(2)})]^{b_{\beta _i}b_{\beta _j}}}=
\langle \exp \left [ \sum _{s,\alpha} \int _{\gamma _s}d\q\, X_{\alpha}^{(s)}(\theta) J_{\alpha}^{(s)}(\theta) \right ] \rangle = \nonumber \\
&&  = \langle \exp \left [ \sum _{i=1}^N a_{\alpha _i} X_{\alpha _i}^{(1)}(\theta_i^{(1)})- \sum _{j=1}^M b_{\beta _j} X_{\beta _j}^{(2)}(-\theta_j^{(2)}) \right ] \rangle \, .
\ea
Therefore, the integrands in (\ref {Whep}) can be factorised into products of functions of the integration variables $\theta_i^{(1)}$, $\theta_j^{(2)}$,
\ba
{\cal W}_{hep}&=&\sum _{N,M=0}^{\infty}\frac{(-1)^{N}}{N!} \langle \int _{\gamma _1}\prod _{i=1}^N \frac{d\theta_i^{(1)}}{2\pi}\left [ \sum _{\alpha} \mu _{\alpha}(\theta_i^{(1)}) \sum_{a_\alpha=1}^\infty\frac{\left (-e^{-\tau _1 E_{\alpha}(\theta_i^{(1)}) +i\sigma _1 p_{\alpha} (\theta_i^{(1)}) +i\phi _1(2-\alpha ) +X^{(1)}_{\alpha}(\theta_i^{(1)})} \right )^{a_\alpha}}{(a_\alpha)^2} \right ] \times \nonumber \\
&\times & \frac{(-1)^{M}}{M!}\int _{\gamma _2}\prod _{j=1}^M \frac{d\theta_j^{(2)}}{2\pi}\left [ \sum _{\beta} \mu _{\beta}(\theta_j^{(2)}) \sum_{b_\beta=1}^\infty \frac{\left (-e^{-\tau _2 E_{\beta}(\theta_j^{(2)}) - i\sigma _2 p_{\beta} (\theta_j^{(2)}) +i\phi _2(2-\beta ) -X^{(2)}_{\beta}(\theta_j^{(2)})} \right )^{b_\beta}}{(b_\beta)^2}\right ] \rangle \nn \, , \\
\ea
so that the sums over $a_{\alpha _i}$ and $b_{\beta _j}$ can be easily performed. Reminding the definition of dilogarithm
$\textrm{Li}_2(z)=\sum_{k=1}^\infty\frac{z^k}{k^2}$, we get
\ba
{\cal W}_{hep}&=& \sum _{N,M=0}^{\infty} \frac{(-1)^{N}}{N!} \langle \int _{\gamma _1}\prod _{i=1}^N \frac{d\theta_i}{2\pi} \sum _{\alpha} \mu _{\alpha}(\theta_i) \textrm{Li}_2 \left (-e^{-\tau _1 E_{\alpha}(\theta_i) +i\sigma _1 p_{\alpha} (\theta_i) +i\phi _1(2-\alpha ) +X^{(1)}_{\alpha}(\theta_i)} \right )  \times \nonumber \\
&\times & \frac{(-1)^{M}}{M!}\int _{\gamma _2}\prod _{j=1}^M \frac{d\theta_j}{2\pi} \sum _{\beta} \mu _{\beta}(\theta_j) \textrm{Li}_2 \left (-e^{-\tau _2 E_{\beta}(\theta_j) - i\sigma _2 p_{\beta} (\theta_j) +i\phi _2(2-\beta ) -X^{(2)}_{\beta}(\theta_j)} \right ) \rangle = \nonumber \\
&=& \langle \exp \left [ - \int _{\gamma _1}\frac{d\theta}{2\pi} \left [ \sum  _{\alpha} \mu _{\alpha}(\theta) \textrm{Li}_2 \left (-e^{-\tau _1 E_{\alpha}(\theta) +i\sigma _1 p_{\alpha} (\theta) +i\phi _1(2-\alpha ) +X^{(1)}_{\alpha}(\theta)} \right ) \right ] \right ] \times \nonumber \\
&\times & \exp \left [ - \int _{\gamma _2}\frac{d\theta}{2\pi} \left [ \sum  _{\beta} \mu _{\beta}(\theta) \textrm{Li}_2 \left (-e^{-\tau _2 E_{\beta}(\theta) -i\sigma _2 p_{\beta} (\theta) +i\phi _2(2-\beta) -X^{(2)}_{\beta}(\theta)} \right ) \right ] \right ] \rangle \ : \nn\\
\ea
this expression allows us to associate, in the strong coupling regime, the heptagonal Wilson loop to a partition function
\be
{\cal W}_{hep}=\int  \prod  _{\alpha , \beta =1}^3 {\cal D}X^{(1)}_{\alpha } {\cal D}X^{(2)}_{\beta } e^{-S[X^{(1)},X^{(2)}]} \, ,
\label {WXX}
\ee
with an effective action
\ba
S[X^{(1)},X^{(2)}]&=&\frac{1}{2} \sum _{s,s',\alpha ,\beta} \int _{\gamma _{s}} d\theta \int _{\gamma _{s'}} d\theta '  X^{(s)}_{\alpha}(\theta ) T^{(s,s')}_{\alpha, \beta}(\theta , \theta ') X^{(s')}_{\beta}(\theta ') + \nonumber \\
&+& \sum _{s,\alpha} \int _{\gamma _{s}} \frac{d\theta}{2\pi} \mu _{\alpha}(\theta) \textrm{Li}_2 \left ( -e^{-\tau _s E_{\alpha}(\theta) + i(-1)^{s+1}\sigma _s p _{\alpha}(\theta) +i\phi _s (2-\alpha)+ (-1)^{s+1}X^{(s)}_{\alpha}(\theta )} \right ) \, , \label {SXX}
\ea
which scales as $S[X^{(1)},X^{(2)}]\sim\sqrt{\lambda}$. This latter observation entitles us to apply, as customary, the saddle point technique: imposing that the action (\ref{SXX}) be at an extremum leads us to a set of integral 'equations of motion' describing the classical configuration for the fields $X^{(s)}_{\alpha}$ :
\ba
&& X^{(s)}_{\alpha}(\theta)+ \sum _{s'=1}^2 \sum _{\alpha '=1}^3 (-1)^{s'} \times \label {Xeq}\\
&\times & \int _{\gamma _{s'}}\frac{d\theta '}{2\pi} \mu _{\alpha '}(\theta ') G^{(s,s')}_{\alpha, \alpha '}(\theta , \theta ')
\ln \left [ 1+e^{-\tau _{s'} E_{\alpha '}(\theta ' ) +i(-1)^{s'-1}\sigma _{s'} p_{\alpha '}(\theta ')+i\phi _{s'}(2-\alpha ') +(-1)^{s'+1}X^{(s')}_{\alpha '}(\theta ')} \right ]=0 \nonumber \, .
\ea
A fruitful redefinition of the fields in terms of 'pseudoenergies'
\be
\varepsilon ^{(s)}_{\alpha} \left ( \theta -i\hat \varphi _s +i\frac{\pi}{4}b_s \right ) = \tau _s E_{\alpha }(\theta ) -i (-1)^{s-1}\sigma _s p_{\alpha }(\theta ) +(-1)^s X^{(s)}_{\alpha }(\theta ) \, ,
\ee
recasts the equations of motion (\ref {Xeq}) into a shape that, once taken into account the relations in Appendix \ref{appA3}, makes manifest the agreement with the classical string result summarized in (\ref{tbaeps1})-(\ref{tbaeps3})
\ba\label{TBAhep}
&&\varepsilon ^{(s)}_{\alpha } ( \theta -i\hat \varphi _s ) = \tau _s E_{\alpha} \left (\theta -i\frac{\pi}{4}b_s \right ) -i (-1)^{s-1}\sigma _s p_{\alpha }\left (\theta - i\frac{\pi}{4}b_s \right )  - \nonumber \\
&& - \sum _{s'=1}^{2} \sum _{\alpha '=1}^3(-1)^{s+s'}\int _{\textrm{Im} \theta '=\hat \varphi _{s'}} d\theta ' \frac{\mu _{\alpha '}\left (\theta '-i\frac{\pi}{4}b_{s'} \right )}{2\pi} G^{(s,s')}_{\alpha , \alpha '}\left (\theta -i\frac{\pi}{4}b_s  , \theta '-i\frac{\pi}{4}b_{s'} \right) \times \nonumber \\
 &&\times  \ln \left ( 1+e^{-\varepsilon ^{(s')}_{\alpha '} \left ( \theta '-i\hat \varphi _{s'} \right )}e^{i\phi _{s'}(2-\alpha ')} \right ) \ .
\ea
Finally, the classical contribution to the effective action may be obtained once the classical configuration for the fields $X^{(s)}_{\alpha}$ is imposed by plugging the solution of (\ref{TBAhep}) into (\ref{SXX}): in doing so, the heptagonal Wilson loop takes the form ${\cal W}_{hep}=\exp (-S_c )$, with $S_c$ proportional to the extremal value of the Yang-Yang functional
(\ref{YYc2})\cite{Anope}:
\ba
S_c &=&\frac{1}{2} \sum _{s,s'=1}^2 \sum _{\alpha, \alpha '=1}^3 \int _{\textrm{Im} \theta =\hat \varphi _s} \frac{d\theta}{2\pi}
\int _{\textrm{Im} \theta '=\hat \varphi _{s'}} \frac{d\theta '}{2\pi} (-1)^{s+s'}\mu _{\alpha} \left ( \theta - \frac{i\pi}{4}b_s \right )  \mu _{\alpha '} \left ( \theta ' - \frac{i\pi}{4}b_{s'} \right ) \times \nonumber \\
&\times & G^{(s,s')}_{\alpha , \alpha '}\left (\theta -i\frac{\pi}{4}b_s  , \theta '-i\frac{\pi}{4}b_{s'} \right) \times \nonumber \\
&\times & \ln \left ( 1+e^{-\varepsilon ^{(s)}_{\alpha } \left ( \theta -i\hat \varphi _{s} \right )}e^{i\phi _{s}(2-\alpha )} \right )\ln \left ( 1+e^{-\varepsilon ^{(s')}_{\alpha '} \left ( \theta '-i\hat \varphi _{s'} \right )}e^{i\phi _{s'}(2-\alpha ')} \right ) + \nonumber \\
&+& \sum _{s=1}^2 \sum _{\alpha=1}^3 \int _{\textrm{Im} \theta =\hat \varphi _s} \frac{d\theta}{2\pi} \mu _{\alpha}\left ( \theta - \frac{i\pi}{4}b_s \right ) \textrm{Li}_2 \left ( -e^{-\varepsilon ^{(s)}_{\alpha } \left ( \theta -i\hat \varphi _{s} \right )+i\phi _s (2-\alpha) }\right ) \, . \label {YY-7}
\ea

\subsection{General case}

The procedure to sum the general $n$-edge polygonal Wilson loop goes through the very same steps as the simplest heptagonal case, once some reasonable generalisations are carried out.
In first place, the multi-particle amplitude can be factorised by applying formula (\ref {P-singl}) \cite{BSV1}, so to get
\ba\label {P-prop}
&&    P_{\vec{A}^{(1)}}(0|\theta_1^{(1)},\ldots , \theta_{N^{(1)}}^{(1)}) P_{\vec{A}^{(1)}\vec{A}^{(2)}}(-\theta_{N^{(1)}}^{(1)},\ldots, -\theta_1^{(1)}|\theta_1^{(2)},\ldots, \theta_{N^{(2)}}^{(2)}) \ldots\times  \nonumber \\
&&    \times P_{\vec{A}^{(n-6)}\vec{A}^{(n-5)}}(-\theta_{N^{(n-6)}}^{(n-6)},\ldots, -\theta_1^{(n-6)}|\theta_1^{(n-5)},\ldots, \theta_{N^{(n-5)}}^{(n-5)}) P_{\vec{A}^{(n-5)}}(-\theta_{N^{(n-5)}}^{(n-5)},\ldots, -\theta_1^{(n-5)}|0)= \nn\\
&&    = \frac{\prod \limits _{s=1}^{n-6}\prod \limits _{i^{(s)}=1}^{N^{(s)}} \prod \limits _{i^{(s+1)}=1}^{N^{(s+1)}} P_{A_{i^{(s)}}^{(s)}A_{i^{(s+1)}}^{(s+1)}}(-\theta_{i^{(s)}}^{(s)}|\theta_{i^{(s+1)}}^{(s+1)})}{\prod \limits _{s=1}^{n-5} \prod \limits _{\stackrel {i^{(s)},j^{(s)}=1}{i^{(s)}\not=j^{(s)}}}^{N^{(s)}} P_{A_{i^{(s)}}^{(s)}A_{j^{(s)}}^{(s)}}(\theta_{i^{(s)}}^{(s)}|\theta_{j^{(s)}}^{(s)}) } \, .
\ea
Then, when the strong coupling limit is concerned, the property (\ref {P-strong}) entails:
\ba
&& P_{\vec{A}^{(1)}}(0|\theta_1^{(1)},\ldots , \theta_{N^{(1)}}^{(1)})\ldots P_{\vec{A}^{(n-5)}}(-\theta_{N^{(n-5)}}^{(n-5)},\ldots, -\theta_1^{(n-5)}|0)= \nn\\
&& =\frac{\prod \limits _{s=1}^{n-6}\prod \limits _{i^{(s)}=1}^{N^{(s)}} \prod \limits _{i^{(s+1)}=1}^{N^{(s+1)}} \left [ P_{\alpha ^{(s)}_{i^{(s)}} \alpha ^{(s+1)}_{i^{(s+1)}}}(-\theta_{i^{(s)}}^{(s)}|\theta_{i^{(s+1)}}^{(s+1)})\right ]^{a^{(s)}_{\alpha _i^{(s)}}a^{(s+1)}_{\alpha _i^{(s+1)}}}}{\prod \limits _{s=1}^{n-5} \prod \limits _{i^{(s)},j^{(s)}=1}^{N^{(s)}} \left [ P_{\alpha ^{(s)}_{i^{(s)}}\alpha ^{(s)}_{j^{(s)}} }(\theta_{i^{(s)}}^{(s)}|\theta_{j^{(s)}}^{(s)})\right]^{a^{(s)}_{\alpha _i^{(s)}} a^{(s)}_{\alpha _j^{(s)}}}} \ .
\ea
Mimicking what has been carried out for the heptagon case, we introduce a set of currents $J^{(s)}_{\alpha}$, with support on the straight line $\gamma_s=\{\q|\textrm{Im}\theta  = \hat\varphi_{s}-\frac{\pi}{4}b_s\}$
\be
J^{(s)}_{\alpha}(\theta )=(-1)^{s+1}\sum _{i=1}^{N^{(s)}}a^{(s)}_{\alpha _i^{(s)}}\delta _{\alpha, \alpha _i^{(s)}}\delta \left (\theta+(-1)^s \theta _i^{(s)}\right ) \, , \quad s=1,...,n-5 \, ,
\ee
as well as the propagators $G^{(s,s')}_{\alpha, \beta}(\theta ,\theta ')$, with support on $\gamma _s \times \gamma _{s'}$, defined such that
\ba\label{green-kernel}
G^{(s,s)}_{\alpha, \beta}(\theta ,\theta ') &=& -\ln [P_{\alpha \beta}(\theta |\theta ')P_{\beta \alpha}(\theta '|\theta )] \, , \quad s=1,...,n-5 \ , \nn\\
G^{(s,s+1)}_{\alpha, \beta}(\theta ,\theta ') &=&-\ln P _{\alpha \beta}\left((-1)^s \theta  | (-1)^s \theta ' \right ) \, , \quad s=1,...,n-6 \ , \\
G^{(s,s-1)}_{\alpha, \beta}(\theta ,\theta ') &=& -\ln P _{\alpha \beta}\left ((-1)^s \theta  | (-1)^s \theta ' \right ) \, , \quad s=2,...,n-5 \ , \nn
\ea
all other elements being zero. Therefore, according to the definitions above and recalling (\ref{identity}), the multi-particle amplitude (\ref {P-prop}) can be conveniently reformulated as
\ba
&&   \frac{\prod \limits _{s=1}^{n-6}\prod \limits _{i^{(s)}=1}^{N^{(s)}} \prod \limits _{i^{(s+1)}=1}^{N^{(s+1)}} \left [ P_{\alpha ^{(s)}_{i^{(s)}} \alpha ^{(s+1)}_{i^{(s+1)}}}(-\theta_{i^{(s)}}^{(s)}|\theta_{i^{(s+1)}}^{(s+1)})\right ]^{a^{(s)}_{\alpha _i^{(s)}}a^{(s+1)}_{\alpha _i^{(s+1)}}}}{\prod \limits _{s=1}^{n-5} \prod \limits _{i^{(s)},j^{(s)}=1}^{N^{(s)}} \left [ P_{\alpha ^{(s)}_{i^{(s)}}\alpha ^{(s)}_{j^{(s)}} }(\theta_{i^{(s)}}^{(s)}|\theta_{j^{(s)}}^{(s)})\right]^{a^{(s)}_{\alpha _i^{(s)}} a^{(s)}_{\alpha _j^{(s)}}}}= \\
&&    =\exp \left[ \frac{1}{2} \sum _{s,s'=1}^{n-5} \sum_{\alpha^{(s)},\alpha^{(s')}=1}^3 \int _{\gamma _s}d\theta  \int _{\gamma _{s'}}d\theta' J^{(s)}_{\alpha^{(s)}}(\theta )\, G^{(s,s')}_{\alpha^{(s)}, \alpha ^{(s')}}(\theta ,\theta ')\,J^{(s')}_{\alpha ^{(s')}}(\theta ') \right]= \nn\\
&&    =\langle \exp \left [ \sum _{s=1}^{n-5}\sum _{\alpha ^{(s)}=1}^{3}\int _{\gamma _s}d\theta  J^{(s)}_{\alpha ^{(s)}}(\theta ) X ^{(s)}_{\alpha ^{(s)}} (\theta ) \right ] \rangle \nn \, .
\ea
The procedure previously displayed for the heptagon can be adapted with no effort to the general $n$-edge polygon case, so to re-sum the general OPE series at strong coupling, eventually obtaining an expression suitable to be interpreted as a partition function
\begin{flalign}
{\cal W}_{n} &= \langle \exp \left [ -\sum _{s=1}^{n-5}\sum _{\alpha ^{(s)}=1}^3 \int _{\gamma _s}\frac{d\theta}{2\pi} \mu _{\alpha ^{(s)}} (\theta)  \textrm{Li}_2 \left ( -e^{-\tau _s E(\theta) +i\sigma_s p(\theta) +i\phi_s (2-\alpha^{(s)})+ (-1)^{s+1}X^{(s)}_{\alpha^{(s)}}((-1)^{s+1}\theta )} \right ) \right ] \rangle \nn\\
&=\int \prod _{s=1}^{n-5} \prod _{\alpha ^{(s)}=1}^3 {\cal D}X^{(s)}_{\alpha ^{(s)}} e^{-S[X^{(1)}...X^{(n-5)}]} \ ,&&
\end{flalign}
where the effective action (again of order $\sim\sqrt{\lambda}$ for $\lambda\rightarrow\infty$) has been introduced
\ba\label{act}
&& S[X^{(1)}...X^{(n-5)}] = \frac{1}{2} \sum _{s,s'=1}^{n-5} \sum _{\alpha ^{(s)}, \alpha ^{(s')}=1}^3\int _{\gamma _{s}} d\theta \int _{\gamma _{s'}} d\theta '  X^{(s)}_{\alpha ^{(s)}}(\theta ) T^{(s,s')}_{\alpha ^{(s)}, \alpha ^{(s')}}(\theta , \theta ') X^{(s')}_{\alpha ^{(s')}}(\theta ') +  \nn\\
&& +\sum _{s=1}^{n-5} \sum _{\alpha ^{(s)}=1}^3\int _{\gamma _{s}} \frac{d\theta}{2\pi} \mu _{\alpha ^{(s)}}(\theta) \textrm{Li}_2 \left ( -e^{-\tau _s E_{\alpha ^{(s)}}(\theta) +
i(-1)^{s+1}\sigma _s p_{\alpha ^{(s)}}(\theta) +  i \phi _s (2-\alpha ^{(s)})+(-1)^{s+1}X^{(s)}_{\alpha ^{(s)}}(\theta )} \right ) \nn \ .\\
\ea
Again, the minimisation of the functional $S[X^{(1)}...X^{(n-5)}]$ \ (\ref{act}) results in a set of equation of motion for the fields $X^{(s)}_{\alpha ^{(s)}}$:
\ba\label{eq_moto}
&& X^{(s)}_{\alpha ^{(s)}}(\theta)+ \sum _{s'=1}^{n-5} \sum _{\alpha ^{(s')}=1}^3(-1)^{s'} \int _{\gamma _{s'}}\frac{d\theta '}{2\pi} \mu _{\alpha ^{(s')}}(\theta ') G^{(s,s')}_{\alpha ^{(s)}\alpha ^{(s')}}(\theta , \theta ') \times \\
&\times &
\ln \left [ 1+e^{-\tau _{s'} E_{\alpha ^{(s')}}(\theta ' ) +i(-1)^{s'-1}\sigma _{s'}p_{\alpha ^{(s')}}(\theta ') +i\phi _{s'}(2-\alpha ^{(s')})+(-1)^{s'-1}X^{(s')}_{\alpha ^{(s')}}(\theta ')} \right ]=0 \nonumber \, .
\ea
As explained before, by means of the pseudoenergies
\be
\varepsilon ^{(s)}_{\alpha} \left ( \theta -i\hat \varphi _s +i\frac{\pi}{4}b_s \right ) = \tau _s E_{\alpha }(\theta ) -i (-1)^{s-1}\sigma _s p_{\alpha }(\theta ) +(-1)^s X^{(s)}_{\alpha }(\theta ) \, ,
\ee
the equations of motion (\ref{eq_moto}) closely resemble some TBA equations
\ba \label{epsbsv}
&&\varepsilon ^{(s)}_{\alpha ^{(s)}} ( \theta -i\hat \varphi _s ) = \tau _s E_{\alpha^{(s)} } \left (\theta -i\frac{\pi}{4}b_s \right ) -i (-1)^{s-1}\sigma _s p_{\alpha^{(s)} }\left (\theta - i\frac{\pi}{4}b_s \right )  - \nonumber \\
&& - \sum _{s'=1}^{n-5} \sum _{\alpha ^{(s')}=1}^3(-1)^{s+s'}\int _{\textrm{Im} \theta '=\hat \varphi _{s'}} d\theta ' \frac{\mu _{\alpha ^{(s')}}\left (\theta '-i\frac{\pi}{4}b_{s'} \right )}{2\pi} G^{(s,s')}_{\alpha ^{(s)}, \alpha ^{(s')}}\left (\theta -i\frac{\pi}{4}b_s  , \theta '-i\frac{\pi}{4}b_{s'} \right) \times \nonumber \\
 &&\times  \ln \left ( 1+e^{-\varepsilon ^{(s')}_{\alpha ^{(s')}} \left ( \theta '-i\hat \varphi _{s'} \right )}e^{i\phi _{s'}(2-\alpha^{(s')})} \right ) \ ,
\ea
although, if we perform the apparently obvious identification $\epsilon_{a,s}=\we^{(s)}_a$, the equations of motion (\ref{epsbsv}) do not seem to actually coincide with the TBA equations retrieved by \cite{TBuA}-\cite{Anope}, though: in fact, unlike the cases already examined, \textit{i.e.} the hexagon and the heptagon, a further step is still required before this task could be achieved, as in next section we will explain.\\
At last, formula (\ref {YY-7}) for the heptagon can be generalised too, since, as previously done, the solutions of the equations (\ref{epsbsv}), suitably plugged into (\ref{act}) lead us to the expression for the classical (extremal) value $S_c$ of the action for the $n$-edge polygonal Wilson loop,
\ba\label{YY-n}
S_c&=&\frac{1}{2} \sum _{s,s'=1}^{n-5} \sum _{\alpha, \alpha '=1}^3 \int _{\textrm{Im} \theta =\hat \varphi _s} \frac{d\theta}{2\pi}
\int _{\textrm{Im} \theta '=\hat \varphi _{s'}} \frac{d\theta '}{2\pi} (-1)^{s+s'}\mu _{\alpha} \left ( \theta - \frac{i\pi}{4}b_s \right )  \mu _{\alpha '} \left ( \theta ' - \frac{i\pi}{4}b_{s'} \right ) \times \nonumber \\
&\times & G^{(s,s')}_{\alpha , \alpha '}\left (\theta -i\frac{\pi}{4}b_s  , \theta '-i\frac{\pi}{4}b_{s'} \right) \times \nonumber \\
&\times & \ln \left ( 1+e^{-\varepsilon ^{(s)}_{\alpha } \left ( \theta -i\hat \varphi _{s} \right )}e^{i\phi _{s}(2-\alpha )} \right )\ln \left ( 1+e^{-\varepsilon ^{(s')}_{\alpha '} \left ( \theta '-i\hat \varphi _{s'} \right )}e^{i\phi _{s'}(2-\alpha ')} \right ) + \nonumber \\
&+& \sum _{s=1}^{n-5}\sum _{\alpha=1}^3 \int _{\textrm{Im} \theta =\hat \varphi _s} \frac{d\theta}{2\pi} \mu _{\alpha}\left ( \theta - \frac{i\pi}{4}b_s \right ) \textrm{Li}_2 \left ( -e^{-\varepsilon ^{(s)}_{\alpha } \left ( \theta -i\hat \varphi _{s} \right )+i\phi _s (2-\alpha) }\right ) \ ,
\ea
such that ${\cal W}_{n}=\exp (-S_c)$. Before comparing (\ref{YY-n}) with the Yang-Yang functional provided by \cite{Anope}, a redefinition of the pseudoenergies $\varepsilon ^{(s)}$ is required, as Section \ref{comparison} below will explain.

\section{Comparisons}\label{comparison}
\setcounter{equation}{0}

A last issue remains still unsolved from the previous Section \ref{predictions}: indeed, the equations of motion (\ref{eq_moto}) and the extremal action (\ref{YY-n}) for the general $n$-edge polygonal Wilson loop do not seem to coincide with the TBA equations (\ref{tbaeps1})-(\ref{tbaeps3}) and extremal Yang-Yang functional (\ref{YYc2}) provided by \cite{TBuA}-\cite{Anope}, unlike the hexagon \cite{FPR2} and the heptagon, which instead manifest the desired agreement. We are now going to show, step by step, how to smooth out this apparent discrepancy, by means of a simple redefinition of the pseudoenergies and a re-parametrisation of the cross ratios.

Let us begin with the latter manipulation. For the cross ratios $\sigma_s$ and $\tau_s$, the simplest choice of a parametrisation in term of the $y_{a,s}$ turns out to work properly
\be
\ln y_{2,s}=-2\tau _s \, , \quad  \ln\frac{y_{2,s}}{y_{1,s}y_{3,s}}=2\sigma _s \label {cross1} \ ,
\ee
so that ${\cal E}_s(\theta )$ is led to the form
\be
{\cal E}_s(\theta )= -2\tau _s \cosh \theta + 2i (-1)^{s-1}\sigma _s \sinh \theta \ .
\ee
Conversely for the last set of cross ratios, the $\phi_s$, a more peculiar identification, depending on the value of the index $s$, will prove itself the most suitable, that is
\be
e^{i\phi _{s}}=\sqrt{\frac{y_{1,s}}{y_{3,s}}} \  , \quad s=4k+1, 4k+2 \ ;
\ e^{i\phi _{s}}=\sqrt{\frac{y_{3,s}}{y_{1,s}}} \  , \quad s=4k+3, 4k+4 \label {cross2} \ .
\ee
Speaking about the measure and propagators appearing in (\ref{eq_moto}) and (\ref{YY-n}), it is straightforward to realise that
\be\label {mu-sh}
\mu _{\alpha} \left (\theta - \frac{i\pi}{4}b_s \right )= \frac{\sqrt{\lambda}}{2\pi}\frac{2}{\sinh ^2 \left [2\theta -\frac{i\pi}{2}b_{\alpha + s} \right ]} \ ,
\ee
whereas a strict correspondence can be established between the components of the propagator $G^{(s,s')}_{\alpha ^{(s)}, \alpha ^{(s')}}$ (\ref {green-kernel}) and the kernels $\tilde K$ (\ref {tildeK}), as may be observed by inspecting Appendix \ref{appA3}. When coping with the pseudoenergies instead, the simplest identification $\epsilon_{a,s}=\we^{(s)}_a$ does not seem to reproduce the expected results, as already pointed out. On the contrary, if we hold the equality $\epsilon_{2,s}=\we^{(s)}_2$ valid $\forall s$ for the index $a=2$, but we 'intertwine' the pseudoenergies for the the values $a=1,3$, according to the values of $s$ as follows ($k \geq 0$ is an integer)
\ba\label{identificazione_eps}
&& \epsilon_{1,4k+1}=\we^{(4k+1)}_1 		\qquad\quad  \epsilon_{3,4k+1}=\we^{(4k+1)}_3 \nn\\
&& \epsilon_{1,4k+2}=\we^{(4k+2)}_1 		\qquad\quad  \epsilon_{3,4k+2}=\we^{(4k+2)}_3 \label{ridefinizione_eps} \\
&& \epsilon_{1,4k+3}=\we^{(4k+3)}_3 		\qquad\quad  \epsilon_{3,4k+3}=\we^{(4k+3)}_1 \nn\\
&& \epsilon_{1,4k+4}=\we^{(4k+4)}_3 		\qquad\quad  \epsilon_{3,4k+4}=\we^{(4k+4)}_1 \nn \, ,
\ea
then one can show that equations (\ref{epsbsv}) do in fact coincide with (\ref{tbaeps1})-(\ref{tbaeps3}). On the same footing, the extremal action (\ref {YY-n}) matches the extremal Yang-Yang functional (\ref {YYc2}) when the identifications (\ref{identificazione_eps}) are taken into account, so that eventually the equality
\be
S_c=\frac{\sqrt{\lambda}}{2\pi} YY_{c} \
\ee
holds.

\section{Y-system}\label{Y-sys}
\setcounter{equation}{0}

In the previous sections we showed how to match the TBA equations (\ref{tbaeps1},\ref{tbaeps2},\ref{tbaeps3}) to the relations (\ref{epsbsv}), arising from the minimisation of the action built out of the re-summation of the OPE series. The interest now turns to the formulation, directly from the equations (\ref{epsbsv}), of a set of functional equations, the so called $Y$-system \cite{Y-system_Zam}, in order to eventually show the agreement with the results by \cite{YSA}.
This effort should be intended as a first attempt towards a deeper understanding of an unusual feature of the scattering amplitude $Y$-system
\cite{YSA}
\ba\label{YSA}
&& \frac{Y^-_{3,s}Y^+_{1,s}}{Y_{2,s}}=\frac{(1+Y_{3,s+1})(1+Y_{1,s-1})}{1+Y_{2,s}} \nn\\
&& \frac{Y^+_{2,s}Y^-_{2,s}}{Y_{1,s}Y_{3,s}}=\frac{(1+Y_{2,s-1})(1+Y_{2,s+1})}{(1+Y_{1,s})(1+Y_{3,s})}\\
&& \frac{Y^-_{1,s}Y^+_{3,s}}{Y_{2,s}}=\frac{(1+Y_{1,s+1})(1+Y_{3,s-1})}{1+Y_{2,s}}  \nn \ ,
\ea
namely its crossed nature, which means the simultaneous presence in the left hand side of the first and third equations (\ref{YSA}) of two different nodes $Y_{a,s}$: in fact, just a few examples \cite{crossed1} \cite{crossed2} \cite{FFPT} are known to exhibit this variant from the customary form of $Y$-systems \cite{Y-system_Zam}.

As a first step, we point out that the equations of motion (\ref{epsbsv}) can be recast into a form resembling (\ref{hatTBA1})-(\ref{hatTBA2}), by introducing the functions $\hat W_{a,s}$, according to the relations
\ba
\we^{(s)}_{1}(\theta -i \hat \varphi _s)&=&-\ln \hat W_{1,s}\left (\theta - \frac{i\pi}{4}b_{s+1} \right ) +i\phi_s \\
\we^{(s)}_{3}(\theta -i \hat \varphi _s)&=&-\ln \hat W_{3,s}\left (\theta - \frac{i\pi}{4}b_{s+1} \right ) -i\phi_s \\
\we^{(s)}_{2}(\theta -i \hat \varphi _s)&=&-\ln \hat W_{2,s}\left (\theta - \frac{i\pi}{4}b_{s} \right ) \ ,
\ea
and also, by mimicking the definition (\ref{hatY})
\be
\hat W_{\alpha,s}(\theta )= W_{\alpha,s} \left ( \theta - \frac{i\pi}{4} b_{\alpha+s+1} \right ) \ ,
\ee
so that, using the relations listed in Appendix \ref{appA3}, the equations (\ref{epsbsv}) turn to
\ba\label{TBA_hat1}
&& \ln \hat W_{2,s} (\theta )-{\cal E}_s(\theta)=-\int_{\textrm{Im} \theta ' =\varphi _{s}} d\theta' \biggl[\tilde K_2^{(s)}\left(\theta, \theta' + \frac{i\pi}{4}b_{s}\right)\Lambda^+_s (\theta') + \nonumber \\
&+& 2\tilde K_1\left(\theta,\theta' + \frac{i\pi}{4}b_{s+1}\right)\Lambda^0_s(\theta')\biggr] +  \int_{\textrm{Im} \theta ' =\varphi _{s-1}} d\theta ' \biggl[\tilde K_1\left(\theta, \theta '+ \frac{i\pi}{4}b_{s+1}\right) \Lambda^+_{s-1}(\theta ') + \nonumber \\
&+& \tilde K_2^{(s)}\left(\theta, \theta' + \frac{i\pi}{4}b_{s}\right) \Lambda^0_{s-1}(\theta')\biggr]  + \int_{\textrm{Im} \theta ' =\varphi _{s+1}} d\theta ' \biggl[\tilde K_1\left(\theta, \theta '+ \frac{i\pi}{4}b_{s+1}\right) \Lambda^+_{s+1}(\theta ') + \nonumber \\
&+& \tilde K_2^{(s)}\left(\theta, \theta' + \frac{i\pi}{4}b_{s}\right) \Lambda^0_{s+1}(\theta')\biggr] \, ,
\ea
\ba
&& \ln \hat W_{1,s} (\theta )+ \ln \hat W_{3,s} (\theta )-\sqrt{2}{\cal E}_s\left (\theta + \frac{i\pi}{4}(-1)^{s+1} \right ) = -\int_{\textrm{Im} \theta ' =\varphi _{s}} d\theta' \biggl[2\tilde K_2^{(s)}\left(\theta, \theta' + \frac{i\pi}{4}b_{s+1}\right)\Lambda^0_s (\theta') + \nonumber \\
&+& 2\tilde K_1\left(\theta,\theta' + \frac{i\pi}{4}b_{s}\right)\Lambda^+_s(\theta')\biggr] +  \int_{\textrm{Im} \theta ' =\varphi _{s-1}} d\theta ' \biggl[\tilde K^{(s)}_2\left(\theta, \theta '+ \frac{i\pi}{4}b_{s+1}\right) \Lambda^+_{s-1}(\theta ') + \nonumber \\
&+& 2\tilde K_1\left(\theta, \theta' + \frac{i\pi}{4}b_{s}\right) \Lambda^0_{s-1}(\theta')\biggr] + \int_{\textrm{Im} \theta ' =\varphi _{s+1}} d\theta ' \biggl[\tilde K^{(s)}_2\left(\theta, \theta '+ \frac{i\pi}{4}b_{s+1}\right) \Lambda^+_{s+1}(\theta ') + \nonumber \\
&+& 2\tilde K_1\left(\theta, \theta' + \frac{i\pi}{4}b_{s}\right) \Lambda^0_{s+1}(\theta')\biggr] \, ,
\ea

\ba\label{TBA_hat3}
\ln \hat W_{1,s} (\theta )- \ln \hat W_{3,s} (\theta ) -2i\phi_s
=(-1)^{s+1}\Biggl[\int_{\textrm{Im} \theta ' =\varphi _{s-1}} d\theta' \tilde K_3 \left(\theta , \theta'+ \frac{i\pi}{4}b_{s+1}\right) \Lambda^-_{s-1}(\theta') + \nonumber \\
+ \int_{\textrm{Im} \theta ' =\varphi _{s+1}} d\theta' \tilde K_3 \left(\theta , \theta'+ \frac{i\pi}{4}b_{s+1}\right) \Lambda^-_{s+1}(\theta')\Biggr] \, ,
\ea
where the functions $\Lambda_s^{\pm}(\q)$ and $\Lambda^0_s(\q)$ stand for
\ba
\Lambda^+_{s}(\theta)=\ln(1+W_{1,s}(\theta))(1+W_{3,s}(\theta)) \, , \quad \Lambda^0_{s}(\theta)=\ln(1+W_{2,s}(\theta)) \, , \quad \Lambda^-_{s}(\theta)=\ln\frac{(1+W_{1,s}(\theta))}{(1+W_{3,s}(\theta))} \ .
\ea
The functions $\hat W_{a,s}$ in $\theta=0$ are related to the parameters $\tau_s$, $\sigma_s$ and $\phi_s$ through a set of equations analogous to (\ref{cross1}) and (\ref{cross2}): they are

\be
\ln \hat W_{2,s}(0)=-2\tau _s \, , \quad  \ln\frac{\hat W_{2,s}(0)}{\hat W_{1,s}(0)\hat W_{3,s}(0)}=2\sigma _s  \, , \quad
e^{i\phi _{s}}=\sqrt{\frac{\hat W_{1,s}(0)}{\hat W_{3,s}(0)}}  \, .
\ee
Moreover, equations (\ref{TBA_hat1})-(\ref{TBA_hat3}) may be further rewritten in terms of relativistic kernels $K_i$ and parameters
$m_s, C_s, \varphi _s$ appearing in equations (\ref{eq1})-(\ref{eq3}):
\ba\label{TBA_rel1}
\ln W_{2,s}(\theta)&=& - |m_s| \sqrt{2} \cosh (\theta - i \varphi _s) -\int_{\textrm{Im} \theta ' =\varphi _{s}} d\theta'\biggl[K_2(\theta-\theta')
\Lambda^+_{s}(\theta) + \nonumber \\
&+& 2K_1(\theta-\theta')\Lambda^0_{s}(\theta')\biggr] + \int_{\textrm{Im} \theta ' =\varphi _{s-1}} d\theta'\biggl[K_2(\theta-\theta')
\Lambda^0_{s-1}(\theta') + \nonumber \\
&+& K_1(\theta-\theta')\Lambda^+_{s-1}(\theta')\biggr] + \int_{\textrm{Im} \theta ' =\varphi _{s+1}} d\theta'\biggl[K_2(\theta-\theta')
\Lambda^0_{s+1}(\theta') + \nonumber \\
&+& K_1(\theta-\theta')\Lambda^+_{s-1}(\theta')\biggr] \, , \\
\ln W_{1,s}(\theta)&=& - |m_s| \cosh (\theta - i \varphi _s) -C_s \left(\sin \frac{\pi s}{2} -\cos \frac{\pi s}{2}\right ) -\int_{\textrm{Im} \theta ' =\varphi _{s}} d\theta'\biggl[K_2(\theta-\theta')
\Lambda^0_{s}(\theta') + \nonumber \\
&+& K_1(\theta-\theta')\Lambda^+_{s}(\theta')\biggr] + \int_{\textrm{Im} \theta ' =\varphi _{s-1}}
d\theta'\biggl[K_1(\theta-\theta')\Lambda^0_{s-1}(\theta') + \nonumber \\
&+& \frac{1}{2}K_2(\theta-\theta')\Lambda^+_{s-1}(\theta') + (-1)^{s+1}\frac{1}{2}K_3(\theta -\theta')\Lambda^-_{s-1}(\theta')\biggr] + \nonumber \\
&+&  \int_{\textrm{Im} \theta ' =\varphi _{s+1}} d\theta'\biggl[K_1(\theta-\theta')\Lambda^0_{s+1}(\theta') + \frac{1}{2}K_2(\theta-\theta')\Lambda^+_{s+1}(\theta') + \nonumber \\
&+& (-1)^{s+1}\frac{1}{2}K_3(\theta -\theta')\Lambda^-_{s+1}(\theta')\biggr] \, ,
\ea
\ba\label{TBA_rel3}
\ln W_{3,s}(\theta)&=& - |m_s| \cosh (\theta - i \varphi _s) +C_s \left(\sin \frac{\pi s}{2} -\cos \frac{\pi s}{2}\right ) -\int_{\textrm{Im} \theta ' =\varphi _{s}} d\theta'\biggl[K_2(\theta-\theta')
\Lambda^0_{s}(\theta') + \nonumber \\
&+& K_1(\theta-\theta')\Lambda^+_{s}(\theta')\biggr] + \int_{\textrm{Im} \theta ' =\varphi _{s-1}}
d\theta'\biggl[K_1(\theta-\theta')\Lambda^0_{s-1}(\theta') + \nonumber \\
&+& \frac{1}{2}K_2(\theta-\theta')\Lambda^+_{s-1}(\theta') -(-1)^{s+1} \frac{1}{2}K_3(\theta -\theta')\Lambda^-_{s-1}(\theta')\biggr] + \nonumber \\
&+&  \int_{\textrm{Im} \theta ' =\varphi _{s+1}} d\theta'\biggl[K_1(\theta-\theta')\Lambda^0_{s+1}(\theta') + \frac{1}{2}K_2(\theta-\theta')\Lambda^+_{s+1}(\theta') - \nonumber \\
&-& (-1)^{s+1}\frac{1}{2}K_3(\theta -\theta')\Lambda^-_{s+1}(\theta')\biggr] \ .
\ea
It should be pointed out that all the differences between the set of equations (\ref{TBA_rel1})-(\ref{TBA_rel3}) and (\ref{eq1})-(\ref{eq3}) are signs multiplying the kernel $K_3$ and the constant $C_s$.

Now that equations (\ref{TBA_rel1})-(\ref{TBA_rel3}) are available, the task of formulating the relative $Y$-system is easily achieved, by taking into account the bootstrap relations (\ref{boot_rel}). Unpleasantly, it turns out that the result explicitly depends on the parity of the label $s$: if $s$ is odd one has
\ba\label{YsysOdd}
\mbox{[$s$ odd]:}&& \nn\\
&& \frac{W^-_{3,s}W^+_{1,s}}{W_{2,s}}=\frac{(1+W_{3,s+1})(1+W_{3,s-1})}{1+Y_{2,s}} \nn\\
&&\frac{W^+_{2,s}W^-_{2,s}}{W_{1,s}W_{3,s}}=\frac{(1+W_{2,s-1})(1+W_{2,s+1})}{(1+W_{1,s})(1+W_{3,s})} \\
&& \frac{W^-_{1,s}W^+_{3,s}}{W_{2,s}}=\frac{(1+W_{1,s+1})(1+W_{1,s-1})}{1+W_{2,s}} , \nn
\ea
whereas even values of $s$ lead to
\ba\label{YsysEven}
\mbox{[$s$ even]:}&& \nn\\
&& \frac{W^-_{3,s}W^+_{1,s}}{W_{2,s}}=\frac{(1+W_{1,s+1})(1+W_{1,s-1})}{1+W_{2,s}}  \nn\\
&&\frac{W^+_{2,s}W^-_{2,s}}{W_{1,s}Y_{3,s}}=\frac{(1+W_{2,s-1})(1+W_{2,s+1})}{(1+W_{1,s})(1+W_{3,s})} \\
&& \frac{W^-_{1,s}W^+_{3,s}}{W_{2,s}}=\frac{(1+W_{3,s+1})(1+W_{3,s-1})}{1+W_{2,s}} \nn \ .
\ea
Although apparently (\ref{YsysOdd}),(\ref{YsysEven}) differ from the $Y$-system provided in \cite{YSA}, the agreement can be easily recovered once an identification analogous to (\ref{identificazione_eps}) is carried out:
\ba\label{ridefinizioneY}
&& Y_{2,s}=W_{2,s}  \\
&& Y_{1,4k+1}=W_{1,4k+1} 		\qquad\quad  Y_{3,4k+1}=W_{3,4k+1} \nn\\
&& Y_{1,4k+2}=W_{1,4k+2} 		\qquad\quad  Y_{3,4k+2}=W_{3,4k+2} \nn\\
&& Y_{1,4k+3}=W_{3,4k+3} 		\qquad\quad  Y_{3,4k+3}=W_{1,4k+3} \nn\\
&& Y_{1,4k+4}=W_{3,4k+4} 		\qquad\quad  Y_{3,4k+4}=W_{1,4k+4} \nn \, .
\ea
In terms of $Y$'s the system (\ref{YsysOdd}),(\ref{YsysEven}) reads exactly the same as (\ref{YSA}).

\medskip

A further way to get the $Y$-system equations (\ref{YsysOdd}),(\ref{YsysEven}) involves the pentagonal amplitudes and bootstrap relations among them, rather than the relativistic kernels. Indeed, the alternative procedure moves the first step directly from the equations of motion (\ref{eq_moto}), which may be rewritten in a shape more suitable for the present purpose
\be\label{eq_moto2}
X^{(s)}_{\alpha}(\theta)+ \sum _{r=1}^{n-5} \sum _{\beta=1}^3(-1)^{r} \int _{\gamma_{r}}\frac{d\theta '}{2\pi} \mu _{\beta}(\theta ') G^{(s,r)}_{\alpha,\beta}(\theta , \theta ')
L_{\beta,r}\left(\q'-\frac{i\pi}{4} b_{r+1}\right)=0 \nonumber \, ,
\ee
where $L_{\alpha,s}(\q)\equiv \ln[1+W_{\alpha,s}(\q)]$.
The $W$-functions are now introduced as
\be
(-1)^s\, X^{(s)}_{\alpha}(\q)=-\ln W_{\alpha,s}\left(\q-\frac{i\pi}{4} b_{s+1}\right)
-\tau_s E_\alpha(\q) -i(-1)^s\sigma_s p_\alpha(\q)+(2-\alpha)\ln\sqrt{\frac{y_{1,s}}{y_{3,s}}} \, ,
\ee
so that the integral equations
\ba
&& \ln\left(\frac{W_{\alpha,s}^+(\q)W_{4-\alpha,s}^-(\q)}{W_{\alpha+1,s}(\q)W_{\alpha-1,s}(\q)}\right)=\sum_{r=1}^{n-5}\sum_{\beta=1}^3(-1)^{r+s}
\int_{\mbox{Im}\q'=\varphi_r} \frac{d\q'}{2\pi}\,\mu_\beta\left (\q'+\frac{i\pi}{4}b_{r+1}\right )L_{\beta,r}(\q')\times \nn \\
&&\times\left[G^{(s,r)}_{\alpha,\beta}\left (\q+\frac{i\pi}{4}b_{s+1}+\frac{i\pi}{4},\q'+\frac{i\pi}{4}b_{r+1}\right )+
G^{(s,r)}_{4-\alpha,\beta}\left (\q+\frac{i\pi}{4}b_{s+1}-\frac{i\pi}{4},\q'+\frac{i\pi}{4}b_{r+1}\right)-\right. \nn\\
&& \left. -G^{(s,r)}_{\alpha+1,\beta}\left (\q+\frac{i\pi}{4}b_{s+1},\q'+\frac{i\pi}{4}b_{r+1}\right)-G^{(s,r)}_{\alpha-1,\beta}\left (\q+\frac{i\pi}{4}b_{s+1},\q'+\frac{i\pi}{4}b_{r+1}\right) \right]
\ea
can be eventually driven to the set of functional equations (\ref{YsysOdd}),(\ref{YsysEven}) by making use of the bootstrap relations (\ref{boot_pentagonal}), involving the elements of the 'Green tensor' (\ref{green-kernel}).

\textbf{Uncrossing the $Y$-system:}\\
Finally, we can manipulate the TBA equations (\ref{TBA_rel1})-(\ref{TBA_rel3}) written in terms of the relativistic kernels $K_1,\,K_2,\,K_3$, in order to turn the unusual (\textit{i.e.} crossed) form of the $Y$-system (\ref{YSA}) into a more conventional one (\textit{i.e.} uncrossed), by means of a set of bootstrap relations (\ref{boot_rel})(\ref{boot_rel2}). Once again the result depends on the value of the $Y$-system column index $s$; indeed, for $s$ odd it is found
\be\label{uncrossed_odd}
  \begin{split}
    W^{++}_{1,s}\,W^{--}_{1,s} &=\frac{(1+W^+_{3,s-1})(1+W^+_{3,s+1})(1+W^-_{1,s+1})(1+W^-_{1,s-1})}
      {(W_{3,s})^2\ (1+\frac{1}{W_{2,s}^+})(1+\frac{1}{W_{2,s}^-})}  \\
    W^{++}_{3,s}\,W^{--}_{3,s} &=\frac{(1+W^+_{1,s-1})(1+W^+_{1,s+1})(1+W^-_{3,s+1})(1+W^-_{3,s-1})}
      {(W_{1,s})^2\ (1+\frac{1}{W_{2,s}^+})(1+\frac{1}{W_{2,s}^-})} \\
    \frac{W^+_{2,s}W^-_{2,s}}{W_{1,s}W_{3,s}} &=\frac{(1+W_{2,s-1})(1+W_{2,s+1})}{(1+W_{1,s})(1+W_{3,s})}  \ ,
  \end{split}
\ee
whereas an even value of $s$ would lead us to
\be\label{uncrossed_even}
  \begin{split}
    W^{++}_{1,s}\,W^{--}_{1,s} &=\frac{(1+W^+_{1,s-1})(1+W^+_{1,s+1})(1+W^-_{3,s+1})(1+W^-_{3,s-1})}
      {(W_{3,s})^2\ (1+\frac{1}{W_{2,s}^+})(1+\frac{1}{W_{2,s}^-})}\\
    W^{++}_{3,s}\,W^{--}_{3,s} &=\frac{(1+W^+_{3,s-1})(1+W^+_{3,s+1})(1+W^-_{1,s+1})(1+W^-_{1,s-1})}
      {(W_{1,s})^2\ (1+\frac{1}{W_{2,s}^+})(1+\frac{1}{W_{2,s}^-})} \\
    \frac{W^+_{2,s}W^-_{2,s}}{W_{1,s}W_{3,s}} &=\frac{(1+W_{2,s-1})(1+W_{2,s+1})}{(1+W_{1,s})(1+W_{3,s})}  \ .
  \end{split}
\ee
On the other hand, the same bootstrap relations (\ref{boot_rel})(\ref{boot_rel2}) may be employed to recast the $Y$-system (\ref{YSA}) to the uncrossed form:
\be
  \begin{split}
    Y^{++}_{1,s}\,Y^{--}_{1,s} &=\frac{(1+Y^+_{1,s-1})(1+Y^-_{1,s+1})(1+Y^+_{3,s+1})(1+Y^-_{3,s-1})}
      {(Y_{3,s})^2\ (1+\frac{1}{Y_{2,s}^+})(1+\frac{1}{Y_{2,s}^-})}\\
    Y^{++}_{3,s}\,Y^{--}_{3,s} &=\frac{(1+Y^+_{3,s-1})(1+Y^-_{3,s+1})(1+Y^+_{1,s+1})(1+Y^-_{1,s-1})}
      {(Y_{1,s})^2\ (1+\frac{1}{Y_{2,s}^+})(1+\frac{1}{Y_{2,s}^-})} \\
    \frac{Y^+_{2,s}Y^-_{2,s}}{Y_{1,s}Y_{3,s}} &=\frac{(1+Y_{2,s-1})(1+Y_{2,s+1})}{(1+Y_{1,s})(1+Y_{3,s})}  \ .
  \end{split}
\ee

\section{Conclusions and outlook}\label{conc}

As extensively illustrated in Section \ref{intro}, {\it Introduction and purposes}, our previous paper \cite{FPR2} has proven the existence, at very large coupling, of the meson and its infinite tower of bound states within the $S$-matrix bootstrap approach (also named fusion in representation theory). Therefore, it has assumed a modification of the spectrum to sum on in the BSV series, as the meson sector contributions revealed their dominance (together with the gluon sector), and yet it has left behind the problem of verifying the actual occurrence of this phenomenon in the BSV series when the coupling is becoming larger and larger. In fact, besides being  relevant in itself, this check may turn out to be useful for further (string) one loop calculations, and also it strengthens, in virtue of its mathematical details, the idea of a liaison with another series coming out of the blue in a different context -- ${\cal N} = 2$ SYM prepotential -- and by other means -- path integral instanton localisation \cite{Nek}. And especially a connection of the strong regime $g \gg 1$ with the Nekrasov-Shatashvili weak $\Omega$-background $\epsilon_2 \sim 0$ \cite{NS,Bourgine2015a,Bourgine2015b}.

Therefore, upon analysing the behaviour of the hexagon BSV series when the coupling grows, we have first shown how the $1$-fermion and $1$-antifermion contribution can be interpreted, in its dominant $g\rightarrow \infty$ part as single particle (mass $=2$) contribution (\textit{cf.} also \cite{BSV3}). This particle is indeed what we dubbed meson in \cite{FPR2}. Successively, we have shown how an intermediate state made up of two fermions and two anti-fermions produces two contributions: the first one may be interpreted as a state of two unbound or {\it free} mesons, the second one as a state of a mass $4$ particle. The latter has received the interpretation of a bound state of two mesons. This pattern perfectly reproduces that used in \cite{FPR2} to re-sum the series to a TBA-like system (with Yang-Yang functional in place of the usual free energy). This mimics somehow the so-called Nekrasov-Shatashvili (NS) limit $\epsilon_2 \to 0$ \cite{NS}, which corresponds to our strong coupling $ g\sim i/ \epsilon_2 \to +\infty$: indeed, in NS limit the leading contribution is an infinite sum over instantons and their bound states which gives rise to a TBA-like equation \cite{Meneghelli, Bourgine2014}. In addition, we have noticed that contributions from unbound fermions are sub-leading; in this sense fermions disappear from the spectrum and give rise to a sort of {\it confinement} at infinite coupling. This phenomenon is instead absent in NS limit, since as we will comment at the end, instantons are elementary ({\it i.e.} non composite) particles.

The effective presence of mesons and their bound states has received evidence in their favour also from this procedure adopted above: supposing their existence, we have extended to a general gluon MHV amplitude the re-summation of the BSV series at large coupling which was performed in \cite {FPR2} only for the six gluon case. Our results agree with the classical string theory minimisation \cite{YSA,Anope}, upon simple redefinition of the pseudoenergies, and thus they reinforce the success of the implementation of the meson and its bound states as particles. Therefore, the latter would be genuine excitations of the {\it classical} GKP string: it would be very interesting to find them as classical solutions. {\it A latere}, we may notice that the only subtlety about the comparison with the string TBA equations \cite{Anope} is that we have needed different integration contours which appear mysteriously (to us) unified in that prolific paper.


Moreover, this methodology could make easier or more transparent the computation of the one-loop contribution by benefitting of the progress concerning the corrections to the Nekrasov-Shatashvili limit \cite{Bourgine2015a,Bourgine2015b}. To make this claim more explicit, it is worth to sketch the form of the Nekrasov instanton partition function for $\m{N}=2\ SU(N_c)$ SYM with $\Omega$-background deformations $\epsilon_1,\epsilon_2$ (with positive imaginary parts) \cite{Nek}:
\be\label{Zinst}
\m{Z} = \sum_{N=0}^{\infty}\frac{\Lambda^N}{N!}\,Z_N=
\sum_{N=0}^\infty\frac{\Lambda^N}{N!}\left(\frac{1}{\epsilon_1\epsilon_2}\right)^N\oint_{\Gamma}\prod_{i=1}^N\frac{d\phi_i}{2\pi i}\,Q(\phi_i)\prod_{i<j}^N K(\phi_i-\phi_j) \ ,
\ee
where $\Lambda$ is the dynamical scale, the potential $Q(\phi)$ is a rational function, the integration contour $\Gamma$ encircles the upper half plane, including the real axis but avoiding possible singularities at the complex infinity, and $K(\phi)$ is the interaction kernel. With respect to $\epsilon_2$, this interaction can be factorised into a short-range part and a long-range one
\be
K(\phi)=\left[1+\frac{\epsilon_2^2}{\phi^2-\epsilon_2^2}\right]e^{\epsilon_2G(\phi)} \, .
\ee
For instance, the two-instanton ($N=2$) contribution to (\ref{Zinst})
\be\label{2inst}
Z_2 = \left(\frac{1}{\epsilon_1\epsilon_2}\right)^2\oint_{\Gamma}\frac{d\phi_1\,d\phi_2 }{(2\pi i)^2}\,Q(\phi_1)Q(\phi_2)\,\left[1+\frac{\epsilon_2^2}{(\phi_1-\phi_2)^2-\epsilon_2^2}\right]e^{\epsilon_2G(\phi_1-\phi_2)}  \ ,\nn
\ee
resembles remarkably the two-meson expression (\ref{4f-fin}), (\ref{splitting}) provided the NS limit $\epsilon_2 \to 0$ on the one side and the strong coupling limit $g\to\infty$ on the other are considered.

Yet, a few important differences stand out, as mentioned in Section \ref{birth}. In particular, the integration contour $\Gamma$ is closed and, remarkably, the potential $Q(\phi)$ and the long range kernel $G(\phi)$ are not analytic within $\Gamma$: in fact, the residues of the former do give corrections to the leading NS limit. On the contrary, the integration path $\m{C}_s$ here is open, and the measure $\hat{\mu}_M(u)$ and the kernel $\displaystyle\frac{1}{P_{reg}^{(MM)}(u|v)P_{reg}^{(MM)}(v|u)}$ in (\ref{4f-fin}) (somehow playing the role of the potential $Q(\phi)$ and the long-range part $e^{\epsilon_2G(\phi_1-\phi_2)}$ respectively) are analytic in the small fermion sheet. Therefore, in order to get a close circuit, the path $\m{C}_s$ needs to be rearranged as a closed curve $\m{C}_{HM}$ plus an interval $\m{I}$ (see Figure \ref{Cfigura2}): therefore, in the strong coupling expansion, some sub-leading corrections precisely arise from the integrations on $\m{I}$ (instead of the closed $\Gamma$). A further prominent difference between the Nekrasov partition function and the hexagonal Wilson loop concerns the physical objects involved. Instantons in the former case, as opposed to mesons: in fact, the latter are not elementary particles, as they arise as bound states of fermions and antifermions. In this respect, the contributions due to free (\textit{i.e.} unbound) fermions (and antifermions) are already neglected in (\ref{4f-fin}) as subdominant. Nevertheless, they must be taken into account when looking for the one-loop corrections to the classical string behaviour\footnote{We point out, for instance, the recent interesting papers \cite{BB1,BB2} towards this direction.}.

\medskip
{\bf Acknowledgements}

It is a pleasure to thank B. Basso, A. Belitsky, L. Bianchi, M. S. Bianchi, J.-E. Bourgine, A. Cavagli\`{a}, L. Dixon, R. I. Nepomechie, G. Papathanasiou, Y. Satoh, A. Sever, R. Tateo and P. Vieira for useful discussions. Hospitality (of DF) at Scuola Normale Superiore (Pisa) is kindly acknowledged. This project was  partially supported by INFN grants GAST and FTECP, the UniTo-SanPaolo research  grant Nr TO-Call3-2012-0088 {\it ``Modern Applications of String Theory'' (MAST)}, the ESF Network {\it ``Holographic methods for strongly coupled systems'' (HoloGrav)} (09-RNP-092 (PESC)) and the MPNS--COST Action MP1210.

\appendix

\section{P functions} \label {functions}
\setcounter{equation}{0}

\subsection{All couplings}\label{allcouplings}
In this appendix, we review some useful all coupling formul\ae\ regarding fermions and mesons (although the latter are real (not virtual) only for infinite coupling), by recalling results and notations from\ \cite{FPR1}. In the first place, the fermionic scattering phases can be written as
\ba\label{Sff}
&& -i\ln S^{(ff)}(u,v)=-i\ln S^{(f\bar f)}(u,v)=
\int_{-\infty}^{+\infty}\frac{dw}{2\pi}\,\chi_H(w,u)\frac{d}{dw}\chi_H(w,v)- \nn\\
&&\qquad -\int_{-\infty}^{+\infty}\frac{dw}{2\pi}\frac{dz}{2\pi}\,\chi_H(w,u)\frac{d^2}{dwdz}\Theta(w,z)\,\chi_H(z,v)   \, ,
\ea
where $\Theta (u,v)$ is connected to the scattering factor $S^{(ss)}$ between two scalars (holes)
\be
{S}^{(ss)}(u,v)= -\textrm{exp} [ -i\Theta (u,v) ] \nn
\ee
and
\be
\chi _H (u,v) = -i \ln \left ( \frac{1-\frac{x^f(v)}{x^+(u)}}{1-\frac{x^f(v)}{x^-(u)}} \right ) \, ,
\ee
upon defining on the two sheets
\be
x(u)=\frac{u}{2}\left (1+\sqrt{1-\frac{4g^2}{u^2}} \right ) \, ,
 \quad x^f(u)=\frac{g^2}{x(u)}=\frac{u}{2}\left (1-\sqrt{1-\frac{4g^2}{u^2}} \right ) \, ,
\ee
with $x^{\pm}(u)=x(u\pm\frac{i}{2})$. We can notice how the fermion-fermion $S$ matrix can be formulated in alternative shapes, as for instance
\be
-i\ln S^{(ff)}(u,v)= - \int _{-\infty}^{+\infty}\frac{dw}{2\pi}\,\chi_H(w,u)\frac{d}{dw} F^{(f)}(w,v) = \sum _{n=0}^{+\infty} \frac{(x^f(u))^{n+1}}{2g^2} Q_{n+2}^{(f)} (v) \, ,
 \ee
involving $F^{(f)}$ (which satisfies the equation (2.64) of \cite {FPR2}), or the $(n+2)$-th charge $Q^{(f)}_{n+2}(v)$ for a small fermion with rapidity $v$. In \cite{BSV3} the procedure is provided to obtain the mirror $S$-matrices involving fermions and antifermions; for clarity, we recall here $S^{(\ast ff)}$,  $S^{(\ast f\bar f)}$
\ba
\ln S^{(\ast f\bar f)}(u,v) &=& \ln S^{(\ast \bar f f)}(u,v)=
\ln S^{(Ff)}(u,v)-\ln S^{(s f)}(u-\frac{i}{2},v) \, , \\
\ln S^{(\ast f f)}(u,v) &=& \ln\left(\frac{u-v}{u-v+i}\right)\,+\ln S^{(\ast \bar f f)}(u,v) \label{psipsi} \, ,
\ea
where $S^{(Ff)}$ and $S^{(s f)}$ are scattering factors large fermion-small fermion and scalar-small fermion, respectively. In terms of functions defined in \cite{FPR1} we have
\ba
&& -i\ln S^{(\ast \bar f f)}(u,v)=
-\int_{-\infty}^{+\infty}\frac{dw}{2\pi}\,\left[\chi_F(w,u)+\Phi (w) \right]\frac{d}{dw}\chi_H(w,v)+ \label {S-rep}\\
&& \qquad +\int_{-\infty}^{+\infty}\frac{dw}{2\pi}\frac{dz}{2\pi}\,\left[\chi_F(w,u)+\Phi (w) \right]\frac{d^2}{dwdz}\Theta(w,z)\,\chi_H(z,v)
-\chi_H(u-\frac{i}{2},v)+ \nn\\
&& \qquad +\int_{-\infty}^{+\infty}\frac{dw}{2\pi}\,\frac{d}{dw}\Theta(u-\frac{i}{2},w)\,\chi_H(w,v) \ , \nn
\ea
with
\be
\chi _F(u,v)+\Phi (u) = i\ln \frac{x^+(u)-x(v)}{x(v)-x^-(u)}+i\ln \left ( - \frac{x^-(u)}{x^+(u)} \right )  \, .
\ee
Eventually, the fermionic pentagonal amplitudes are expressed as \cite{BSV3}
\ba
\left[P^{(ff)}(u|v)\right]^2 &=& \frac{f_{ff}(u,v)}{(u-v)^2}\,\frac{S^{(ff)}(u,v)}{S^{(\ast \bar ff)}(u,v)} \, , \\
\left[P^{(f\bar f)}(u|v)\right]^2 &=& - \frac{S^{(ff)}(u,v)}{f_{ff}(u,v) S^{(\ast \bar ff)}(u,v)} \, ,
\ea
where it has been defined
\be
f_{ff}(u,v)=\frac{x^f(u)x^f(v)}{g^{2}} -1 \,  .
\ee
Indeed, the first square above, $\left[P^{(ff)}(u|v)\right]^2$, can also be obtained from its representation (38) in \cite{BSV3} upon use of (\ref{psipsi}); while our second square $\left[P^{(f\bar f)}(u|v)\right]^2$ differs by a sign from that in \cite{BSV3}. Now, some comments are due on the analyticity of these two pentagonal amplitudes. In fact, it is reasonable to think that $S^{(\ast f\bar f)}$, $S^{(f f)}$  and $f_{ff}$ have no zeroes nor poles in the small fermion sheet because of their representations above: in this direction, the analogous forms (257) and (143) of \cite {BEL1407} could be the most suited ones. Besides, physically a pole in $S^{(\ast f\bar f)}$ or $S^{(f f)}$ would signal the presence of a new particle. This absence of poles and zeroes of $S^{(\ast f\bar f)}$ and $S^{(ff)}$ entails that $P^{(f\bar f)}$ is analytic and non-zero in the small fermion sheet, as stated by \cite {BSV3} and exploited in several computations therein\footnote{However, it seems to us that
a rigorous proof of the analytic-non-zero property of $S^{(\ast f\bar f)}$ and $S^{(ff)}$ is still lacking.}. Instead, $P^{(ff)}(u|v)$ has (no zeroes and) only one single pole for coinciding rapidities $u=v$, and this fact allows one to show the existence of a fermionic measure $\mu_f(u)\neq0$:
\be
\lim_{v\rightarrow u}\, (v-u) P^{(ff)}(u|v)=\frac{i}{\mu_f(u)} \ .
\ee

The $S$-matrix describing the scattering of two mesons can thus be obtained by fusing the $S$-matrices of their fermion constituents, so to obtain the {\it formal} (all couplings) relation
\be\label{ff->M}
S^{(MM)}(u,v)=\frac{u-v+i}{u-v-i} S^{(ff)}(u+i,v+i)S^{(ff)}(u-i,v+i)S^{(ff)}(u+i,v-i)S^{(ff)}(u-i,v-i) \, .
\ee
Inspired by this relation, 
we make the proposal for the {\it formal} (all couplings) $P$ factor between two mesons
\be\label{PMM}
P^{(MM)}(u|v)=-(u-v)(u-v+i) P^{(ff)}(u+i|v+i) P^{(ff)}(u-i|v-i) P^{(f\bar f)}(u-i|v+i)
P^{(f\bar f)}(u+i|v-i) \, ,
\ee
which evidently satisfies the Watson's equation
\be
S^{(MM)}(u,v)=\frac{P^{(MM)}(u|v)}{P^{(MM)}(v|u)} \ .
\ee
and the other axioms. In terms of (fermionic) $S$-matrices, it can be re-written as
\ba\label{PiMM}
P^{(MM)}(u|v)&=& \sqrt{\frac{f_{ff}(u+i,v+i) f_{ff}(u-i,v-i)}{f_{ff}(u-i,v+i) f_{ff}(u+i,v-i)}} \times \label {P(MM)}\\
&\times & \sqrt{\frac{S^{(ff)}(u+i,v+i) S^{(ff)}(u-i,v-i)S^{(ff)}(u+i,v-i)S^{(ff)}(u-i,v+i)}{S^{(\ast ff)}(u+i,v+i) S^{(\ast ff)}(u-i,v-i)S^{(\ast f\bar f)}(u+i,v-i)S^{(\ast f\bar f)}(u-i,v+i)}} \nonumber \, .
\ea
Thanks to (\ref{psipsi}) the denominator of the second square root of (\ref {P(MM)}) may be expressed in terms of $S^{(\ast f\bar f)}$ only:
\ba
P^{(MM)}(u|v)&=&   \frac{u-v+i}{u-v} \sqrt{\frac{f_{ff}(u+i,v+i) f_{ff}(u-i,v-i)}{f_{ff}(u-i,v+i) f_{ff}(u+i,v-i)}} \times \label {P(MM)2}\\
&\times & \sqrt{\frac{S^{(ff)}(u+i,v+i) S^{(ff)}(u-i,v-i)S^{(ff)}(u+i,v-i)S^{(ff)}(u-i,v+i)}{S^{(\ast f\bar f)}(u+i,v+i) S^{(\ast f\bar f)}(u-i,v-i)S^{(\ast f\bar f)}(u+i,v-i)S^{(\ast f\bar f)}(u-i,v+i)}}= \nn\\
&=&  \frac{u-v+i}{u-v} P^{(MM)}_{reg}(u|v)  \nonumber \, .
\ea
This expression shows that $P^{(MM)}(u|v)$ has only one simple pole for $u=v$ and one simple zero for $u=v-i$: no other poles or zeroes are present in the small sheet. As a consequence $P^{(MM)}_{reg}$ does not contain any pole or zero.

Our proposal (\ref {P(MM)2}) is confirmed by analysing the strong coupling limit.
Indeed, when $\lambda \rightarrow \infty$ and we turn to the hyperbolic variables $u=2g\coth 2\theta$, $v=2g\coth 2\theta '$, formula (\ref {P(MM)2}) reproduces (10.15) of \cite {FPR2}

\be
P^{(MM)}(\theta | \theta')=1-\frac{i}{2g}\frac{\sinh 2\theta \sinh 2\theta '}{ \sinh (2\theta -2\theta ')}
\sqrt{2}\cosh \left (\theta -\theta ' -i\frac{\pi}{4} \right ) +O(1/g^2)  \, ,  \label {KMMstrongsmall}
\ee
which we have obtained by solving the axioms for the mesonic pentagonal amplitudes directly at strong coupling.
To prove that, one can use (C.42) of \cite {FPR2} for the fermion-fermion phase and (C.50) of the same paper for the mirror $S$-matrix between a fermion and an antifermion, which read

\be
S^{(ff)}(\theta,\theta') =
\exp\left\{-\frac{i}{4g}\frac{\sinh2\theta \sinh2\theta'}{\sinh(2\theta-2\theta')}\,(\cosh(\theta-\theta')-1)
+O\left(\frac{1}{g^2}\right)\right\} \, , \label {Sfftheta}
\ee

\be
S^{(\ast f\bar f)}(\theta ,\theta') =
\exp\left\{\frac{\sinh2\theta \sinh2\theta'}{8g\cosh(\theta-\theta')}\,
+O\left(\frac{1}{g^2}\right)\right\} \, ,
\label {Sfbarfmir}
\ee
together with the (all couplings) expression of the factor $f_{ff}(u,v)$ in terms of hyperbolic rapidities
\be
f_{ff}(\theta , \theta ')= \tanh \theta \tanh \theta ' -1 \, .
\ee

The behaviour (\ref{KMMstrongsmall}) can be also obtained from formula (\ref{PMM}), after using the fermionic pentagonal transitions at strong coupling, formul\ae\ (10.18) of \cite{FPR2}
\ba
&&[P^{(ff)}(\theta | \theta ')]^2 = -\frac{\sinh \theta \sinh \theta ' \sinh 2\theta \sinh 2\theta '}{2g^2 \sinh (\theta -\theta ') \sinh (2\theta -2\theta ')} \Bigl [ 1+\frac{i}{4g}\frac{\sinh 2\theta \sinh 2\theta '}{\sinh (2\theta -2\theta ')} \left ( 1-\cosh (\theta -\theta ')+i\sinh (\theta -\theta ')\right ) \Bigr ] \nn \\
&&[P^{(f\bar f)}(\theta |\theta ')]^2 = \frac{\cosh \theta \cosh \theta '}{\cosh (\theta -\theta ')}
\Bigl [ 1+\frac{i}{4g}\frac{\sinh 2\theta \sinh 2\theta '}{\sinh (2\theta  -2\theta ')} \left ( 1-\cosh (\theta -\theta ')+i\sinh (\theta -\theta ')\right ) \Bigr ]
\label {Pfbarf} \, .
\ea

\subsection{Pentagonal amplitudes at strong coupling}\label{appA2}

Here is a collection of the functions $P_{\alpha \beta}(\theta | \theta ')$:
\ba
P_{11}(\theta | \theta ') &=& P_{33}(\theta | \theta ')=1+\frac{i\pi}{\sqrt{\lambda}} \frac{\cosh 2\theta \cosh 2\theta '}{\sinh (2\theta -2\theta ')}\left [1+\cosh (\theta -\theta ')-i\sinh (\theta -\theta ') \right ]=  \label {Pgg} \\
&=& 1+ \frac{2\pi}{\sqrt{\lambda}}\,K^{(gg)}(\q,\q')=1+ \frac{2\pi}{\sqrt{\lambda}}\,K^{(\bar g\bar g)}(\q,\q') \nn \, , \ea
\ba
P_{13}(\theta | \theta ') &=& P_{31}(\theta | \theta ')=1+\frac{i\pi}{\sqrt{\lambda}} \frac{\cosh 2\theta \cosh 2\theta '}{\sinh (2\theta -2\theta ')}\left [-1+\cosh (\theta -\theta ')-i\sinh (\theta -\theta ') \right ] =  \label {Pgbarg} \\
&=& 1+ \frac{2\pi}{\sqrt{\lambda}}\,K^{(g\bar g)}(\q,\q')=1+ \frac{2\pi}{\sqrt{\lambda}}\,K^{(\bar g g)}(\q,\q') \nn
\, ,
\ea
\ba
P_{22}(\theta | \theta ') &=& 1-\frac{2\pi}{\sqrt{\lambda}}\frac{i \sinh 2\theta \sinh 2\theta '}{ \sinh (2\theta -2\theta ')} \sqrt{2}\cosh \left (\theta -\theta ' -i\frac{\pi}{4} \right ) =  \label {PmMM} \\
&=& 1+ \frac{2\pi}{\sqrt{\lambda}}\,K^{(MM)}(\q,\q') \nn \, ,
\ea
\ba
P_{21}(\theta | \theta ') &=& P_{23}(\theta | \theta ')=1+\frac{2\pi}{\sqrt{\lambda}}\frac{\sinh 2\theta \cosh 2\theta '}{\sqrt{2} \cosh (2\theta -2\theta ')}[ \sinh (\theta -\theta ')+i \cosh (\theta -\theta ') ] =\\
&=& 1+ \frac{2\pi}{\sqrt{\lambda}}\,K^{(Mg)}(\q,\q')=1+ \frac{2\pi}{\sqrt{\lambda}}\,K^{(M\bar g)}(\q,\q') \nn \, ,
\ea
\ba
P_{12}(\theta | \theta ') &=& P_{32}(\theta | \theta ')=1+\frac{2\pi}{\sqrt{\lambda}} \frac{\sinh 2\theta ' \cosh 2\theta }{\sqrt{2} \cosh (2\theta ' -2\theta )}[ \sinh (\theta ' -\theta )-i \cosh (\theta '-\theta ) ] =\\
&=& 1+ \frac{2\pi}{\sqrt{\lambda}}\,K^{(gM)}(\q,\q')=1+ \frac{2\pi}{\sqrt{\lambda}}\,K^{(\bar gM)}(\q,\q') \nn \, .
\ea

\subsection{Relations between kernels}\label{appA3}

From the definition (\ref{green-kernel}) and the formul{\ae} in Appendix \ref{appA2}, at strong coupling we have the following equalities

\be
G^{(s,s)}_{1,1}(\theta,\theta')=G^{(s,s)}_{1,3}(\theta,\theta')=G^{(s,s)}_{3,1}(\theta,\theta')=
G^{(s,s)}_{3,3}(\theta,\theta') \, ,
\ee
\be
G^{(s,s)}_{1,2}(\theta,\theta')=G^{(s,s)}_{3,2}(\theta,\theta') \, , \quad G^{(s,s)}_{2,1}(\theta,\theta')=G^{(s,s)}_{2,3}(\theta,\theta') \, ,
\ee
\be
G^{(s,s+1)}_{1,2}(\theta,\theta')=G^{(s,s+1)}_{3,2}(\theta,\theta') \, , \quad G^{(s,s+1)}_{2,1}(\theta,\theta')=G^{(s,s+1)}_{2,3}(\theta,\theta') \, ,
\ee
\be
G^{(s,s+1)}_{1,1}(\theta,\theta')=G^{(s,s+1)}_{3,3}(\theta,\theta') \, , \quad G^{(s,s+1)}_{1,3}(\theta,\theta')=G^{(s,s+1)}_{3,1}(\theta,\theta') \, ,
\ee
in addition with the obvious ones $G^{(s,s+1)}_{\alpha,\beta}(\theta,\theta')=G^{(s,s-1)}_{\alpha,\beta}(\theta,\theta')$.

We list also some of the relations between the kernels $G$ and the tilded kernels $\tilde K$:
\be
\frac{\mu _2(\theta ')}{2\pi} G^{(s,s)}_{2,2}(\theta , \theta ') =-2 \tilde K_1 (\theta , \theta ') \, ,
\ee
\be
\frac{\mu _2(\theta ')}{2\pi} G^{(s,s+1)}_{2,2}(\theta , \theta ') =
-\tilde K_2^{(s)}(\theta , \theta ') \, ,
\ee
\begin{equation}
\frac{\mu_1(\theta')}{2\pi}G^{(s,s)}_{2,1}(\theta,\theta')=-\tilde K^{(s)}_2\left(\theta,\theta'+i\frac{\pi}{4}(-1)^s\right) \, ,
\end{equation}

\begin{equation}
\frac{\mu_1(\theta')}{2\pi}G^{(s,s+1)}_{2,1}(\theta,\theta')=-\tilde K_1\left(\theta,\theta'-i\frac{\pi}{4}(-1)^s\right) \, ,
\end{equation}


\begin{equation}
\frac{\mu_2(\theta')}{2\pi}G^{(s,s)}_{1,2}(\theta,\theta')=-\tilde K^{(s)}_2\left(\theta-i\frac{\pi}{4}(-1)^s,\theta'\right) \, ,
\end{equation}

\begin{equation}
\frac{\mu_1(\theta')}{2\pi}G^{(s,s)}_{1,1}(\theta,\theta')=-\tilde K_1\left(\theta+i\frac{\pi}{4}(-1)^s,\theta'+i\frac{\pi}{4}(-1)^s\right) \, ,
\end{equation}


\begin{equation}
\frac{\mu_2(\theta')}{2\pi}G^{(s,s+1)}_{1,2}(\theta,\theta')=-\tilde K_1\left(\theta+i\frac{\pi}{4}(-1)^s,\theta'\right) \, ,
\end{equation}


\begin{equation}
\frac{\mu_1(\theta')}{2\pi}\left[G^{(s,s+1)}_{1,1}(\theta,\theta')+G^{(s,s+1)}_{3,1}(\theta,\theta')\right]=-\tilde K^{(s)}_2\left(\theta-i\frac{\pi}{4}(-1)^s,\theta'-i\frac{\pi}{4}(-1)^s\right) \, ,
\end{equation}



\begin{equation}
\frac{\mu_1(\theta')}{2\pi}\left[G^{(s,s+1)}_{1,1}(\theta,\theta')-G^{(s,s+1)}_{3,1}(\theta,\theta')\right]=(-1)^s \tilde K_3\left(\theta+i\frac{\pi}{4}(-1)^s,\theta'-i\frac{\pi}{4}(-1)^s\right) \, .
\end{equation}



\section{Bootstrap relations} \label{bootstrap}
\setcounter{equation}{0}

Below, we display a list of bootstrap relations involving the relativistic kernels. We made use of the shorthand notation $K_a(\q^\pm)=K^\pm_a(\q)=K_a(\q \pm\frac{i\pi}{4})$ and also $K_a(\q^{\pm\pm})=K^{\pm\pm}_a(\q)=K_a(\q \pm\frac{i\pi}{2})$.
\ba\label{boot_rel}
&& K_1^{+}+K_1^{-}=K_2 \, , \nn\\
&& K_2^{+}+K_2^{-}=2K_1+\delta(\q) \, ,\\
&& K_3^{-}-K_3^{+}=\delta(\q) \nn \, ,
\ea
\ba\label{boot_rel2}
&& K_1^{++}+K_1^{--}=\delta(\q) \nn \, , \\
&& K_2^{++}+K_2^{--}=\delta(\q^+)+\delta(\q^-) \, , \\
&& K_3^{++}+K_3^{--}=2K_3-\delta(\q^+)+\delta(\q^-) \ . \nn
\ea
The bootstrap relations involving pentagonal amplitudes are:
\ba\label{boot_pentagonal}
&& K^{(MM)}_{sym}\left(\q^{++},\q'^+\right)+ K^{(MM)}_{sym}\left(\q,\q'^+\right)
   -2K^{(gM)}_{sym}\left(\q^+,\q'^+\right) =0 \, , \\
&& K^{(Mg)}_{sym}\left(\q^{++},\q'^+\right)+ K^{(Mg)}_{sym}\left(\q,\q'^+\right)
   -2K^{(gg)}_{sym}\left(\q^+,\q'^+\right) =\pi \sinh^2(2\q)\,\delta(\q-\q') \, , \nn\\
&& K^{(gM)}\left(\q',\q^{++}\right)+ K^{(gM)}\left(\q',\q\right)
   -K^{(gg)}\left(\q',\q^+\right)-K^{(g\bar g)}\left(\q',\q^+\right) =0 \nn \, , \\
&& K^{(MM)}\left(\q',\q^{++}\right)+ K^{(MM)}\left(\q',\q\right)
   -2K^{(Mg)}\left(\q',\q^+\right) = \pi \sinh^2(2\q)\,\delta(\q-\q') \nn \, , \\
&& K^{(gM)}_{sym}\left(\q^{++},\q'^+\right)+ K^{(gM)}_{sym}\left(\q,\q'^+\right)
   -K^{(MM)}_{sym}\left(\q^+,\q'^+\right) = -\pi \cosh^2(2\q)\,\delta(\q-\q') \nn \, , \\
&& K^{(gg)}_{sym}\left(\q^{++},\q'^+\right)+ K^{(gg)}_{sym}\left(\q,\q'^+\right)
   -K^{(Mg)}_{sym}\left(\q^+,\q'^+\right) = 0  \nn \, , \\
&& K^{(Mg)}\left(\q',\q^{++}\right)+ K^{(Mg)}\left(\q',\q\right)
   -K^{(MM)}\left(\q',\q^+\right) = 0  \nn \, , \\
&& K^{(gg)}\left(\q',\q^{++}\right)+ K^{(g\bar g)}\left(\q',\q\right)
   -K^{(gM)}\left(\q',\q^+\right) = 0  \nn \, , \\
&& K^{(g\bar g)}\left(\q',\q^{++}\right)+ K^{(gg)}\left(\q',\q\right)
   -K^{(gM)}\left(\q',\q^+\right) = -\pi \cosh^2(2\q)\,\delta(\q-\q')  \nn \, , \\
&& K^{(Mg)}\left(\q^+,\q'^+\right)+ K^{(Mg)}\left(\q^-,\q'^+\right)
   -K^{(g\bar g)}\left(\q,\q'^+\right)- K^{(gg)}\left(\q,\q'^+\right) = 0 \nn \, , \\
&& K^{(MM)}\left(\q^+,\q'^+\right)+ K^{(MM)}\left(\q^-,\q'^+\right)
   -2K^{(gM)}\left(\q,\q'^+\right) = -\pi \cosh^2(2\q)\,\delta(\q-\q')  \nn \, , \\
&& K^{(Mg)}_{sym}\left(\q^+,\q'\right)+ K^{(Mg)}_{sym}\left(\q^-,\q'\right)
   -2K^{(gg)}_{sym}\left(\q,\q'\right) = -\pi \cosh^2(2\q)\,\delta(\q-\q') \nn \, , \\
&& K^{(MM)}_{sym}\left(\q^+,\q'\right)+ K^{(MM)}_{sym}\left(\q^-,\q'\right)
   -2K^{(gM)}_{sym}\left(\q,\q'\right) = 0 \nn \, , \\
&& K^{(g\bar g)}\left(\q^-,\q'^+\right)+ K^{(gg)}\left(\q^+,\q'^+\right)
   -K^{(Mg)}\left(\q,\q'^+\right) = \pi \sinh^2(2\q)\,\delta(\q-\q') \nn \, , \\
&& K^{(gg)}\left(\q^-,\q'^+\right)+ K^{(g\bar g)}\left(\q^+,\q'^+\right)
   -K^{(Mg)}\left(\q,\q'^+\right) = 0 \nn \, , \\
&& K^{(gM)}\left(\q^-,\q'^+\right)+ K^{(gM)}\left(\q^+,\q'^+\right)
   -K^{(MM)}\left(\q,\q'^+\right) = 0 \nn \, , \\
&& K^{(gg)}_{sym}\left(\q^-,\q'\right)+ K^{(gg)}_{sym}\left(\q^+,\q'\right)
   -K^{(Mg)}_{sym}\left(\q,\q'\right) = 0 \nn \, , \\
&& K^{(gM)}_{sym}\left(\q^-,\q'\right)+ K^{(gM)}_{sym}\left(\q^+,\q'\right)
   -K^{(MM)}_{sym}\left(\q,\q'\right) = \pi\sinh^2(2\q)\, \delta(\q-\q') \nn \, ,
\ea
where the shifts shall be read as $\q^\pm=\q\pm\frac{i\pi}{4}$ and $\q^{\pm\pm}=\q\pm\frac{i\pi}{2}$. In terms of the Green tensor (\ref{green-kernel}), the relations above can be summarised as:
\ba
 G^{(s,s)}_{\alpha,\beta}(\q^{+},\q')+G^{(s,s)}_{4-\alpha,\beta}(\q^{-},\q')
  -G^{(s,s)}_{\alpha+1,\beta}(\q,\q')-G^{(s,s)}_{\alpha-1,\beta}(\q,\q') =
  -2\pi \delta_{\alpha+\beta,odd}\,\frac{\delta(\q-\q')}{\mu_\beta(\q')} \nn \, , \\
 G^{(2k\pm 1,2k)}_{\alpha,\beta}(\q^{++},\q')+G^{(2k\pm 1,2k)}_{4-\alpha,\beta}(\q,\q')
  -G^{(2k\pm 1,2k)}_{\alpha+1,\beta}(\q^+,\q')-G^{(2k\pm 1,2k)}_{\alpha-1,\beta}(\q^+,\q') =
  -2\pi \delta_{4-\alpha,\beta}  \,\frac{\delta(\q-\q')}{\mu_\beta(\q')}  \nn \, , \\
 G^{(2k,2k\pm 1)}_{\alpha,\beta}(\q^{+},\q'^+)+G^{(2k,2k\pm 1)}_{4-\alpha,\beta}(\q^-,\q'^+)
  -G^{(2k,2k\pm 1)}_{\alpha+1,\beta}(\q,\q'^+)-G^{(2k,2k\pm 1)}_{\alpha-1,\beta}(\q,\q'^+) =
  -2\pi \delta_{\alpha,\beta}  \,\frac{\delta(\q-\q')}{\mu_\beta(\q'^+)} \nn \, ,\\
\ea
where $k=1,2,3,\dots$ \,.


\end{document}